\documentclass[preprint2]{proto}
\usepackage{times}
\usepackage{natbib}
\usepackage{float}
\usepackage{graphicx}
\usepackage{pslatex}

\begin{document}
\title{\textbf{\LARGE Jets and Outflows From Star to Cloud:\\
Observations Confront Theory}\\}
\author {\textbf{\large A.\ Frank$^{\rm 1}$, T.P.\ Ray$^{\rm 2}$, S.\ Cabrit$^{\rm 3}$, 
P.\ Hartigan$^{\rm 4}$,
H.G.\ Arce$^{\rm 5}$, F.\ Bacciotti$^{\rm 6}$, 
J.\ Bally$^{\rm 7}$, M.\ Benisty$^{\rm 8}$, \\ 
J.\ Eisl\"offel$^{\rm 9}$, M.\ G\"udel$^{\rm 10}$,
  S.\ Lebedev$^{\rm 11}$, B.\ Nisini$^{\rm 12}$ \& A.\ Raga$^{\rm 13}$}}
\affil{\small\em 1.\ University of Rochester, 2.\ Dublin Institute for Advanced Studes, 3.\ LERMA, Observatoire de Paris, 4.\ Rice University, 5.\ Yale University, 
6.\ Osservatorio Astrofisico di Arcetri, Florence 7.\ University of Colorado at 
Boulder, 8.\ University of Grenoble, 9.\ Th\"uringer Landessternwarte,
10.\ University of Vienna,  11.\ Imperial College 
London, \\ 
12.\ Osservatorio Astronomico di Roma, 13.\ Universidad Nacional Aut\'onoma de M\'exico.}

\begin{abstract}
\baselineskip = 11pt
\leftskip = 0.65in 
\rightskip = 0.65in
\parindent=1pc
{\small In this review we focus on the role jets and outflows play in the star and planet formation process. Our essential question can be posed as follows: are jets/outflows merely an epiphenomenon associated with star formation or do they play an important role in mediating the physics of assembling stars both individually and globally? We address this question by reviewing the current state of observations and their key points of contact with theory. Our review of jet/outflow phenomena is organized into three length-scale domains: Source and Disk Scales ($0.1-10^2$ au) where the connection with protostellar and disk evolution theories is paramount; Envelope Scales ($10^2-10^5$ au) where the chemistry and propagation shed further light on the jet launching process, its variability and its impact on the infalling envelope; Parent Cloud Scales ($10^5-10^6$ au) 
where global momentum injection into cluster/cloud environments become relevant. Issues of feedback are of particular importance on the smallest scales where planet formation regions in a disk may be impacted by the presence of disk winds, irradiation by jet shocks or shielding by the winds. Feedback on envelope scales may determine the final stellar mass (core-to-star efficiency) and envelope dissipation. Feedback also plays an important role on the larger scales with outflows contributing to turbulent support 
within clusters including alteration of cluster star formation efficiencies (feedback 
on larger scales currently appears unlikely). In describing these observations we also look to the future and consider the questions that new facilities such as ALMA and the Jansky 
Array can address. A particularly novel dimension of our review is that we consider
results on jet dynamics from the emerging field of High Energy Density Laboratory Astrophysics (HEDLA). HEDLA is now providing direct insights into the 3-D dynamics 
of fully magnetized, hypersonic, radiative outflows. \\~\\~\\~}


\end{abstract}


\section{\textbf{INTRODUCTION}}

In many ways the discovery that star formation involves {\it outflow} as well as {\it inflow} from gravitational collapse marked the beginning of modern studies of the assembly of stars.  Jets and outflows were the first and most easily observed recognition that the narrative of star formation would include many players and processes beyond the spherical collapse of clouds.  The extraordinary progress made in the study of protostellar jets and outflows since the first discovery of Herbig-Haro (HH) objects (1950s), HH Jets (1980s) and molecular outflows (1980s) also reflects the growing power and sophistication of star formation science. The combination of ever higher resolution observational and computational methods, combined with innovative laboratory experiments, have allowed many aspects of the protostellar outflow problem to be clarified, though as we shall see crucial issues such as the launching process(es) remain debated. 

Hypersonic collimated protostellar mass-loss appears to be a ubiquitous aspect of the star formation process. The observations currently indicate that most, if not all, low and high mass stars produce accretion-powered collimated ejections during their formation. These ejections are traced in two ways (see eg. the excellent PPV observational reviews by \citet{2007prpl.conf..231R}, \citet{2007prpl.conf..245A}, \cite{2007prpl.conf..215B}). 

\begin{figure*}[t]
 \epsscale{1.5}
 \plotone{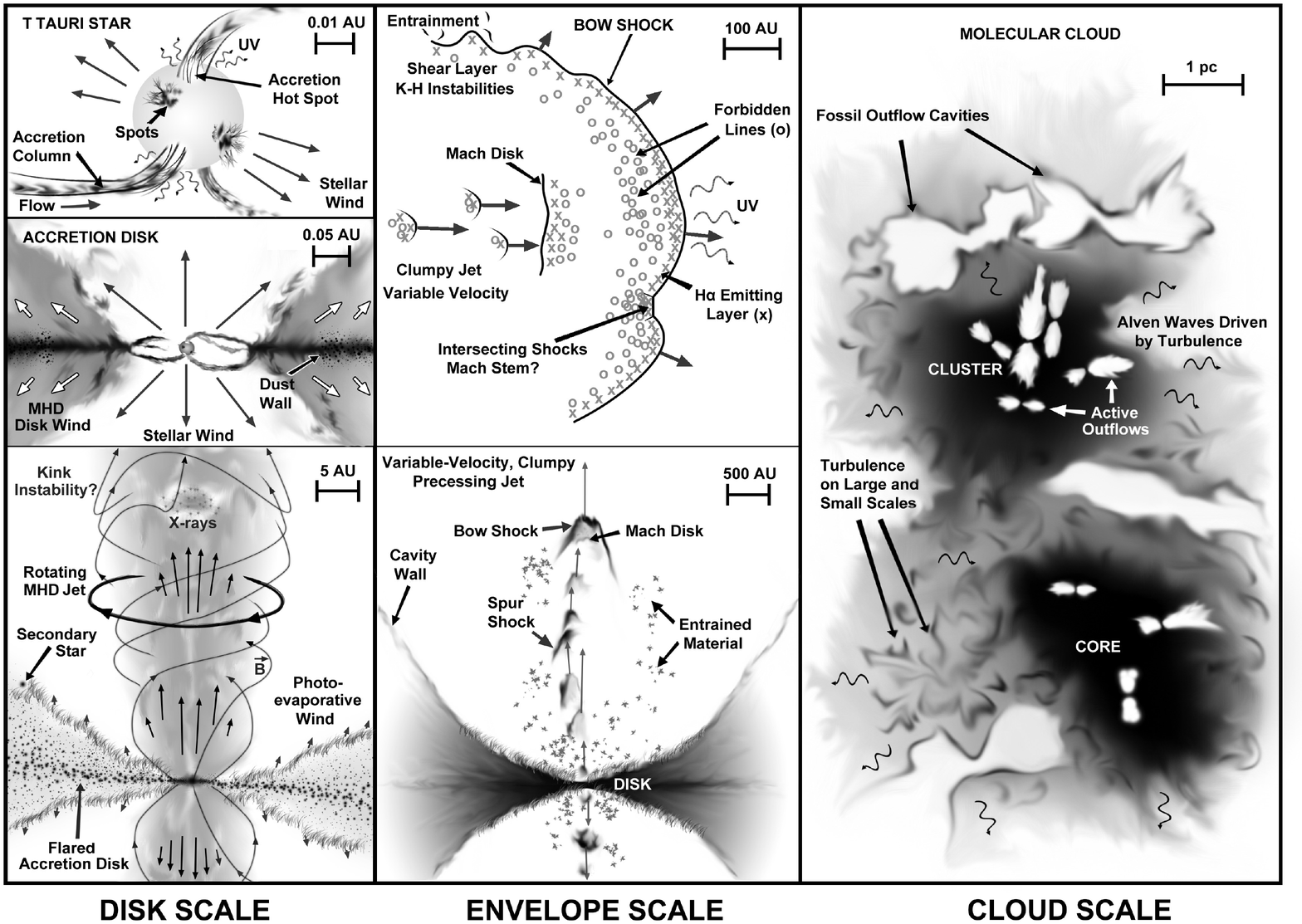}
 \caption{\small A schematic view of jets and outflows across seven orders of magnitude in scale.  Note the presence of the scale bar in each figure as one moves from the physics of launching near the star out to the physics of feedback on cluster and cloud scales. See text for reference to specific processes and classes of objects} 
\label{Hartigan}
\end{figure*}

First there are the narrow, highly-collimated ``jets'' of atomic and/or molecular gas with velocities of order $v \sim 100-1000$ km/s ($v$ increasing with central source mass). These jets are believed to arise through 
magneto-hydrodynamic (MHD) processes in the rotating star-disk system. The other tracer 
are the less collimated more massive ``molecular outflows'' with velocities of order $v \sim 1-30$ km/s which are believed to consist of shells of ambient gas swept-up by the jet bowshock
and a surrounding slower wider-angle component. The fast and dense jet quickly escapes from the protostellar envelope (the still infalling remains of the original ``core'' from which the star formed) and propagates into the surrounding environment to become a ``parsec-scale outflow". The less dense wide-angle wind and the swept-up outflow expand more slowly, carving out a cavity which widens over time into the envelope and the surrounding cloud. Most stars are born in clustered environments where the stellar separation is $< 1\,pc$.  Thus, these large-scale outflows affect the interstellar medium within a cluster and, perhaps, the cloud as a whole. The important elements and processes on each scale from star to cloud are illustrated in Fig.\,\ref{Hartigan}. 

Given their ubiquity and broad range of scales, a central question is whether jets and outflows constitute a mere epiphenomenon of star formation, or whether they are an essential component in the regulation of that process. In particular, winds/outflows are currently invoked to solve several major outstanding issues in star formation: (1) the low star formation efficiency in turbulent clouds (see eg. chapters by Padoan et al., Krumholtz et al.), (2) the systematic shift between the core mass function and the stellar initial mass function, suggesting a  core-to-star efficiency of only 30\% (see eg. chapters by Offner et al., Padoan et al.), (3) the need to efficiently remove angular momentum from the young star and its disk. The former is important to avoid excessive spin-up by accretion and contraction, (cf chapter by Bouvier et al.) while the latter is required to maintain accretion at observed rates, in particular across the ``dead-zone" where MHD turbulence is inefficient (cf. chapter by Turner et al.). Last but not least, protostellar winds may also affect disk evolution and planet formation through disk irradiation or shielding, and enhanced radial mixing of both gas and solids. 

In this chapter we address this central question of outflow feedback on star and planet formation while also reviewing the current state of jet/outflow science. From the description above, it is clear that the degree of feedback will differ according to outflow properties on different spatial scales. The impact on the star and disk will depend on the physics of jet launching and angular momentum extraction (small scales). The impact on core-to-star efficiencies will depend both on the 
intrinsic jet structure and the jet propagation/interaction with surrounding gas. Finally, the impact on global star formation efficiency will depend on the overall momentum injection and on the efficacy of its coupling to cloud turbulence. 

Thus in what follows we review the current state of understanding of protostellar jets and outflows by breaking the chapter into 3 sections through the following division of scales: Star and Disk  (1-10$^{\rm 2}$\,au); Envelope and Parent Clump (10$^{\rm 2}$au - 0.5pc); Clusters and Molecular Clouds (0.5 - 10$^{\rm 2}$\,pc). In each section we review the field {\it and} present new results obtained since the last Protostars and Planets 
meeting.  Where appropriate we also address how new results speak to issues of feedback on star and planet formation. We also attempt to point to ways in which new observing platforms such as ALMA can be expected to influence the field in the near future.  We note that we will focus on outflows from nearby low-mass stars ($<$ 500 pc), which offer the best resolution into the relevant processes. An excellent  review of outflows from high-mass sources was presented in \citet{2007prpl.conf..245A}, which showed that some (but not all) appear as scaled-up versions of the low-mass case \citep[see also e.g.][]{2013A&A...550A..81C,2013ApJ...767...58Z}. ALMA will revolutionize our view of these distant and tightly clustered objects so much that our current understanding will greatly evolve in the next few years. Finally we note that our review includes, for the first time, results on jet dynamics from the emerging field of High Energy Density Laboratory Astrophysics (HEDLA).  These experiments and their theoretical interpretation provide direct insights into the 3-D dynamics of fully magnetized, hypersonic, radiative outflows. 
 



\section{\textbf{SOURCE AND DISK SCALES (1-10$^{\rm 2}$\,au)}}

\subsection{The Accretion-Ejection Link}
\label{Accretion-Outflow}
While the precise origin of jets from young stars is still hotly debated, 
there is general consensus that the launching process involves 
the dynamical interaction of accreted matter with the stellar and/or disk magnetic field. 
However, launch distances depend on the model:  
a few $R_\odot$ for stellar winds, $\simeq 0.05$au for 
those launched at the stellar magnetosphere-disc interface 
\citep[see chapter by Bouvier et al. and][]{2007prpl.conf..261S}, and possibly as far as several au for magneto-centrifugal Disk (D)-winds  \citep[see chapter by Turner et al., and][]{2007prpl.conf..277P}. Unfortunately the projected dimensions on the sky for even the largest proposed launch regions are tens of milliarcseconds. We do not, however, have to spatially resolve the launch zone to at least begin to test various models. Jet properties such as the ejection to accretion ratio, collimation, and velocity structure can be measured on larger scales which then allow mechanisms working on smaller scales to be inferred. That said, we note however that interferometric studies are beginning to resolve the smaller scales directly (see \S\,\ref{Interferometry}).

We first consider the ratio of jet mass-flux to accretion rate. Measuring the mass outflow rate in an atomic jet can be achieved since in principle we know all the necessary quantities from observation. For example through spectroscopy 
and multi-epoch imaging, we can determine both the radial and tangential 
velocity of a jet and hence its true velocity. In addition the jet radius, 
ionisation fraction, electron density and hence total density can be found from a combination of imaging and consideration of various line ratios \citep{2011A&A...527A..13P}. 
One can then calculate $\dot M_{jet} \approx \pi {r_{jet}}^2 \rho_{jet} V_{jet}$. Typical outflow rates are found to be 10$^{-7}$ to 10$^{-9}\, M_{\odot}\,yr^{-1}$ 
for jets from low-mass classical T Tauri stars (CTTSs). As one might suspect, 
higher rates are found for more embedded sources of comparable 
mass \citep{2012A&A...538A..64C}. 
We caution that what we see as a jet in fact consists of a string of shocks with a wide range of conditions. 
The measurements described above represent a bulk average over the shocked gas, unless the knots are well-resolved spatially. Moreover there are a number of methods of measuring the mass-loss rate that give somewhat different values (within a factor 3-10). For a more detailed examination of this 
problem the reader is referred to \citet{2010LNP...793..213D}.

Measuring the accretion rate, ${\dot M}_{acc}$, is also challenging.  
Assuming material is accreted onto the star through magnetospheric accretion \citep{2007prpl.conf..479B}
 from the vicinity of the disk's inner radius $R_{in}$ 
yields the accretion luminosity $L_{acc} \approx GM_\ast {\dot M}_{acc}(1-R_\ast/R_{in})/R_\ast$.
Note that the disk inner radius is often considered to be its co-rotation radius
with the star. 
In the case of CTTSs, this energy is 
mainly observed in the UV-band \citep{2000ApJ...544..927G}, but 
direct observation of this UV excess can be difficult as it may be highly extincted,
particularly in more embedded sources. Fortunately the strength of the UV excess has been found to be 
related to the luminosity of a number of optical and infrared emission lines 
such as H$\alpha$, CaII, Pa$\beta$ and Br$\gamma$  \citep[e.g. ][]{2006A&A...452..245N}, 
which are thought to be mainly produced in the (magnetospheric) accretion 
funnel flow. The relationships between the various line luminosities and the 
UV excess has been tested for objects from young brown dwarfs up to intermediate mass
young stars and has been found to be robust \citep[e.g. ][]{2012A&A...548A..56R}. 

These emission line ``proxies'' can be used to determine the accretion luminosity, and hence accretion 
rates, with a good degree of certainty. The large instantaneous spectral coverage made possible by new instruments such as XSHOOTER on the VLT is particularly well suited to simultaneously cover both accretion and jet line indicators and thus to constrain the ejection/accretion ratio \citep{2013A&A...551A...5E}.  

A number of important caveats must however be raised in considering these methods. First, spectro-astrometric or interferometric studies of certain lines show 
that some portion of their emission must 
arise from the outflow, i.e.\ not all of the line's luminosity can be from magnetospheric 
accretion close to the star \citep[][and  \S\,\ref{Interferometry}]{2009ApJ...691L.106W}. In such cases, the good correlation with UV excess would trace in part the underlying ejection-accretion connection.

Moreover as the accretion does not seem to be uniform, i.e.\ there may be an unevenly spaced number of accretion columns (see Fig.\,\ref{Hartigan}), individual line strengths can vary 
over periods of days with the rotation phase of the star \citep{2012MNRAS.427.1344C}. Accretion can also be intrinsically time-variable on shorter timescales than those probed by forbidden lines in jets (several yrs). Thus time-averaged accretion values should be used when comparisons are made with mass-flux rates derived from such jet tracers. 

Studies of accretion onto YSOs suggest a number of findings that are directly relevant to outflow studies. In particular it is found that:
\begin{description}
\item[$\bullet$] Once the dependence on stellar mass ($\propto M_\star^2$) is taken into account, the accretion rate seems to fall off with time $t$ with an approximate $t^{-1}$ law \citep{2012A&A...538A..64C}. This also seems to be reflected in outflow proxies, with similar ejection/accretion ratios in Class I and Class II sources \citep[e.g.][]{2008A&A...479..503A}.

\item[$\bullet$]Many embedded sources appear to be accreting at instantaneous rates that are far too low to acquire final masses consistent with the initial mass function \citep{2009ApJS..181..321E, 2012A&A...538A..64C}. This suggests accretion and associated outflows may be episodic.

\item[$\bullet$] Typical ratios of jet mass flux to accretion rate for low-mass CTTS are $\simeq$ 10\% \citep[e.g.][]{2007IAUS..243..203C}.
Similar ratios are obtained for jets from 
intermediate-mass T Tauri stars \citep{2009A&A...493.1029A} and
Herbig Ae/Be stars \citep[e.g ][]{2013A&A...551A...5E}. In contrast, 
the lowest mass Class II objects (e.g., young brown dwarfs) show larger ratios \citep{2009ApJ...706.1054W}. This begs the obvious question: do very low mass objects have difficulty 
accreting because of their magnetic ejection configuration ?
\end{description}

If the jet is responsible for extracting excess angular momentum from the accretion disk, then the angular momentum flux in the wind/jet, $\dot J_W$, should equal that to be removed from the accreting flow, $\dot J_{acc}$. Since $\dot J_W \simeq \dot M_W \Omega {r_A}^2$ and 
$\dot J_{acc} \simeq \Omega {r_l}^2\dot M_{acc}$ (where $\Omega$ is the angular velocity at the launch radius ${r_l}$, and ${\rm r_A}$ is the Alfv\'en radius) it follows that $\dot M_W/\dot M_{acc} \simeq (r_l/r_A)^2 = 1/\lambda$, with $\lambda$ defined as the magnetic 
lever arm parameter of the disk wind \citep{1982MNRAS.199..883B}. The observed ejection/accretion ratio of 10\% in CTTS is then consistent with a moderate $\lambda \simeq 10$, while the higher ratio in brown dwarfs would indicate that the Alv\'en radius is much closer to the launch point.

\subsection{The Collimation Zone}
\label{Collimation}

Since PPV, great strides have been made in our understanding of how jets are collimated from both theoretical and observational perspectives. In particular recent images of jets within 100\,au of their source give us clues as to how 
the flows are focused. These observations involve  
both high spatial resolution instrumentation in space, e.g. HST, as well as ground based studies, (e.g. 
various optical/IR AO facilities and mm/radio interferometers). At PPV it was already known 
that optically visible jets from classical T Tauri stars, (i.e.\ Class II sources), begin with wide (10-30 degree) 
opening angles close to the source and are rapidly collimated to within a few degrees in the innermost 50--100\,au
\citep{2007prpl.conf..231R}.  

\begin{figure}[t]
 \epsscale{1.0}
 \plotone{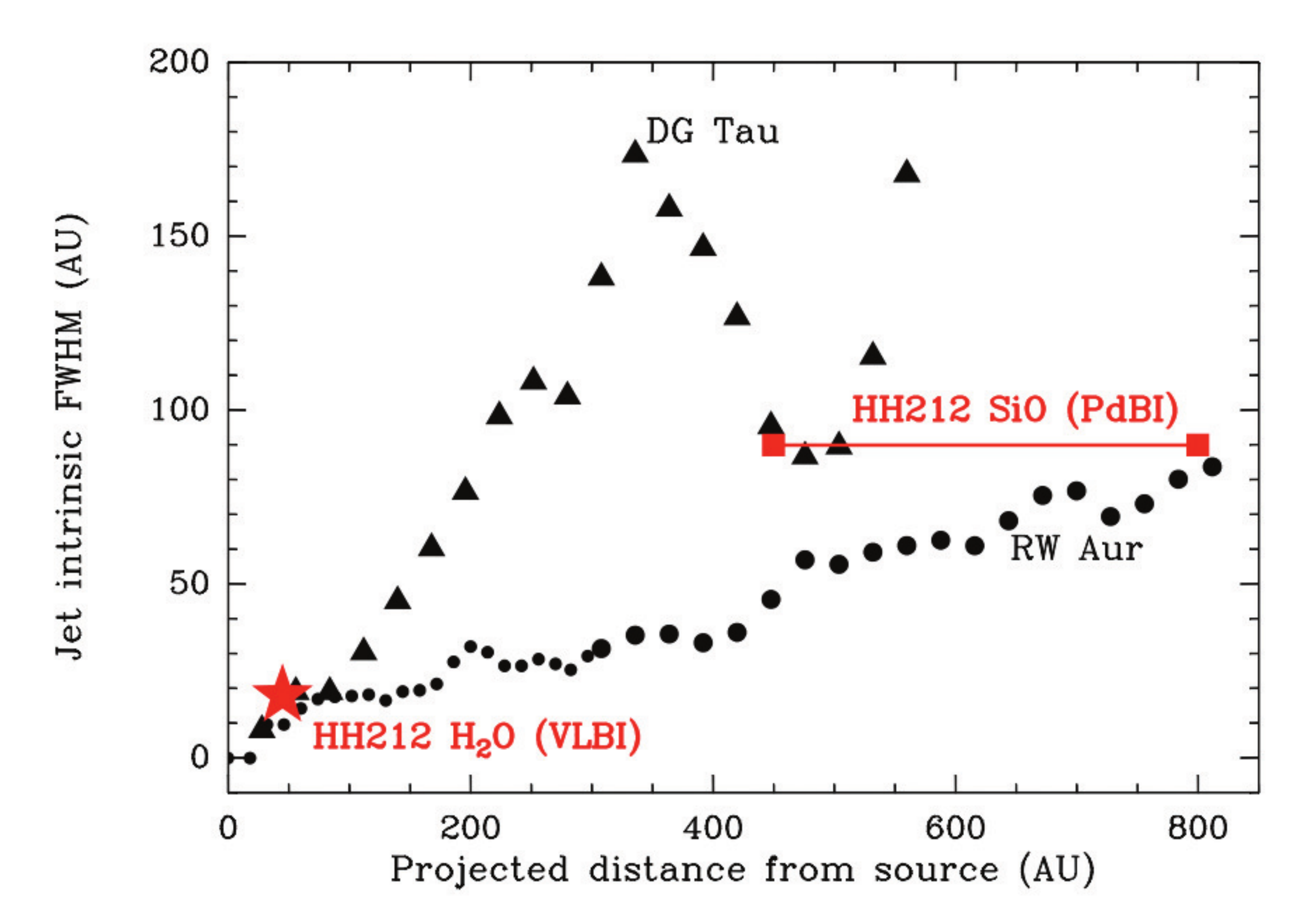}
 \caption{\small A plot showing the width of the jet in HH\,212 as observed in SiO 
 with the PdB Interferometer versus distance from the source in au. A comparison with
 jets from Class II sources (e.g.\ DG\,Tau and RW\,Aur as illustrated) shows that this outflow,
 from an embedded Class 0 source, is collimated on similar scales. From \citet{2007A&A...468L..29C}}
 \label{Cabrit}
\end{figure}

Perhaps the most interesting finding since PPV is that rapid focusing of jets occurs not 
only in the case of Class II sources but also in {\em embedded protostars as well}, 
i.e. \ Class 0 
and Class I sources. Jets from these early phases are difficult to observe optically, due to the 
large amount of dust present, and certainly one cannot trace them optically back to their 
source as in the case of classical T Tauri stars. Nevertheless, their inner regions can be probed 
through molecular tracers such as SiO in the millimeter range, 
or [Fe II] and H$_2$ lines in the near and mid-infrared (see Section~\ref{sec:multiple}).
Using for example the IRAM Plateau de Bure Interferometer (PdBI) with 0\farcs3 resolution, \citet{2007A&A...468L..29C} has shown that the SiO jet from the Class 0 source HH\,212 is collimated on scales similar to jets from 
Class II sources. This suggests that the infalling envelope does not play a major role in focusing the jet.  Thus a more universal
collimation mechanism must be at work at all stages of star formation to produce a directed beam of radius about
15 au on 50 au scales (see Fig.\ref{Cabrit}). The same applies to Class I jets, where both the ionized and molecular jet components show similar opening angles at their base as Class II jets \citep{ 2011A&A...528A...3D}. 
Observations of this type rule out collimation by the ambient thermal pressure \citep{2009pjc..book..247C} and 
instead favour the idea, first proposed by \citet{1988ApJ...332L..41K}, that magnetic fields {\em anchored in the disk} 
force the jet to converge. The required poloidal disk field would be $B_D \simeq 10{\rm mG} ({\dot M}_{\rm w} V_{\rm w} /10^{-6} M_\odot {\rm  km\, s^{-1} \,yr^{-1}})^{0.5}$ where the scaling is for typical CTTS jet parameters.

Thanks to the additional toroidal field that develops in the centrifugal launch process, MHD disk winds could provide the required collimation with an even smaller poloidal disk field \citep{2006A&A...460....1M}.
In this case, we expect distortion of the magnetic field, from a largely poloidal to a largely toroidal geometry,
to begin in the vicinity of the Alfv\'en radius (${\rm r_A} = \sqrt{\lambda} r_l$). The jet however may have 
to traverse many Alfv\'en radii before being effectively focused, since its collimation depends not simply on the magnetic lever arm but also on the poloidal field strength at the disk surface. Several models of truncated MHD disk winds reproduce the PSF-convolved widths of {\em atomic} Class II jets with launch radii in the range 0.1--1 au, despite widely differing magnetic lever arms \citep{2010A&A...516A...6S, 2010ApJ...714.1733S}.

In the future, ALMA and then JWST will be able to carry out even more detailed high spatial resolution 
studies of outflows that should better distinguish between these models. In the interim, the new class of high sensitivity radio interferometers, such as the JVLA and e-MERLIN, are already on stream and show potential 
for measuring the collimation of jets within 10\,au of the source 
\citep{2013MNRAS.tmpL.163A, 2013ApJ...766...53L}.

\subsection{Angular Momentum Transport in Jets}
\label{Angular_Momentum}
Irrespective of the precise nature of the central engine, a basic 
requirement of any complete model is that angular momentum must be 
removed from the accreted material before it can find its final
'resting place' on the star. As matter rains down on the disk from the 
surrounding envelope before being accreted, this process must involve
the disk. As reviewed in the chapter by {\em Turner et al.} in this volume, ordinary particle viscosity is too small to make the {\em horizontal transport} of angular momentum 
from  inner to outer regions of the disk efficient, and additional mechanisms have to be considered. A promising mechanism 
appeared to be the generation of a ``turbulent viscosity''
by  magneto-rotational instabilities, 
the so-called MRI \citep{2006ApJ...652.1020B}. 
Recent studies, however, 
demonstrate that in an initially MRI-unstable disk, the inclusion of a significant vertical magnetic flux, and of ambipolar diffusion coupled with Ohmic dissipation, suppress 
MRI turbulence and instead a powerful magneto-centrifugal wind is generated \citep{2013ApJ...769...76B,Lesur2013}.
Indeed, MHD centrifugal models for jet launching 
\citep{1982MNRAS.199..883B,1983ApJ...274..677P} indicate  
that  protostellar jets can provide a valid solution 
to the angular momentum problem via {\em vertical transport} along the ordered 
component of the strong magnetic field threading the disk. 
In D-wind models this occurs in an extended region where 
the foot-points of the flow are located 
\citep[see e.g.,][]{1997A&A...319..340F,2007prpl.conf..277P}, while in the
X-wind model it is assumed that the wind only extracts the angular momentum from the inner boundary of the disk  
\citep{1990RvMA....3..234C,1994ApJ...429..781S,2009ApJ...692..346F,2013ApJ...768....5C}. Note that in this case the angular momentum ``problem'' is only 
partially resolved as material still has to be transported within the disk to its inner truncation radius.
Finally, MHD stellar winds flowing along open
field lines attached to the star's surface \citep[e.g.][]{1999A&A...348..327S,2005ApJ...632L.135M}, 
and episodic plasmoid ejections by magnetospheric field lines linking the 
star and the disk \citep{2000MNRAS.312..387F,2013A&A...550A..99Z}, 
contribute to the braking of the star (see chapter by Bouvier et al.). In fact such stellar and/or magnetospheric 
winds must be active, at some level, to explain the low observed spin rates of young stars.
Thus several MHD ejection sites probably coexist in young stars, and the difficulty is to determine the relative contribution of each to the observed jets.


A key observational diagnostic to discriminate between these theories is the detection of possible signatures of rotation in protostellar jets. 
The review by \citet{2007prpl.conf..231R} in 
PPV describes the detection, in 5 objects, of   
asymmetric Doppler shifts in emission lines from 
opposite borders of the flow \citep{2000MNRAS.314..241D,2002ApJ...576..222B,2005A&A...432..149W,2004ApJ...604..758C,2007ApJ...663..350C}. 
Since PPV, Doppler shifts have been searched for in many other outflows, in atomic {\em and} 
molecular lines. These studies are very 
demanding, as they require both high angular and spectral resolution, pushing the instrumentation, even on HST, to its limits. 
Possible signatures of rotation, with toroidal and poloidal  
velocities $v_{\phi}$, $v_p$ consistent with magneto-centrifugal acceleration appear in many of the cases studied 
(HL Tau: \citep{2007A&A...470..605M}, HH 26, HH 72: \citep{2008A&A...482..575C}, HH 211: \citep{2007ApJ...670.1188L,2009ApJ...699.1584L}, HH212: \citep{2008ApJ...685.1026L, 2011A&A...526A..40C}, CB 26: \citep{2009A&A...494..147L}, Ori-S6: \citep{2010A&A...510A...2Z},  NGC 1333 IRAS 4A2: \citep{2011ApJ...728L..34C}. 

The collection of the rotation data  allowed for 
the determination of the  specific angular momentum $rv_{\phi}$. 
Assuming an axisymmetric, stationary magneto-centrifugal wind, 
the ratio $rv_{\phi}/{v_p}^2$ gives the location $r_l$ of 
the foot-point in the disk of the sampled streamline, while the product $rv_{\phi}v_p$ gives the magnetic lever arm parameter $\lambda$ \citep[e.g.][]{2003ApJ...590L.107A,2006A&A...453..785F}. Hence this is a powerful tool to discriminate between proposed jet spatial origins and launch models. As shown in Fig.~5 of \citet{2009pjc..book..247C}, the observed signatures when interpreted as steady jet rotation are only consistent with an origin in an extended, warm D-wind, launched from between 0.1 to 3-5 au. {\em The significant implication is that jets and the associated magnetic fields may strongly affect the disk structure in the
region where terrestrial planet form}. The inferred magnetic lever arm parameter is moderate, $\lambda \leq 10$, in line with the mean observed ejection to accretion ratio (see \S~\ref{Accretion-Outflow}).

Note however, that due to limited angular resolution, only the external streamlines of the flow are sampled \citep{Pesenti2004}, and the current measurements cannot exclude the existence of inner  stellar or X-winds. In addition, all the measurements are based on emission lines produced in shocks, that can also self-generate rotational motions \citep{2011ApJ...737...43F}.
Finally, if the jet is observed far from the star, the interaction with the environment can 
hide and confuse rotation signatures.
The primary hypothesis to be tested, however, is the veracity
or otherwise of the rotation interpretation. 
Simulations including an imposed rotation motion were successful in reproducing the observed spectra  
\citep[e.g.] []{2010ApJ...722.1325S}. In contrast, other studies claim that  rotation 
can be mimicked  by e.g. asymmetric shocking against a warped disk \citep{2005A&A...435..125S}, jet
precession \citep{2006A&A...448..231C}, and internal shocks \citep{2011ApJ...737...43F}. Although it is unlikely that these processes apply in all cases, they may contribute to the Doppler shift, confusing the real rotation signature.

From an observational perspective it was found that out of 
examined disks associated with rotating jets, one, 
the   RW\,Aur disk, clearly appeared to counter-rotate with respect to the jet
\citep{2006A&A...452..897C}. Since this result potentially undermined the rotation hypothesis the bipolar jet from RW Aur was observed again after the SM4 repair, twice with STIS in 
UV lines, at an interval of six months \citep{2012ApJ...749..139C}.
The result was again puzzling: the rotation sense for one lobe was in agreement 
with the disk, 
and hence opposite to that measured in the optical
years before. Moreover no signature was detected at that time from the other lobe and, 
after six months, it had disappeared from both lobes. 
Despite these findings, \citet{2012ApJ...759L...1S} has recently
demonstrated that disagreement with the disk rotation can be accommodated 
within the classical magneto-centrifugal theory, as toroidal velocity reversals can occur 
occasionally without violating the total (kinetic plus magnetic) angular momentum conservation. Their simulations also show that the
rotation sense can change in time, thereby accounting for the detected variability.   
Thus it appears that observations are still compatible with the 
jets being a robust mechanism for the extraction of angular momentum from the inner disk. The gain in resolution offered by ALMA and JWST will be crucial to test and confirm this interpretation.

\subsection{Wide-angle structure and blue/red asymmetries}

\begin{figure}[t]
\epsscale{1.0}
\plotone{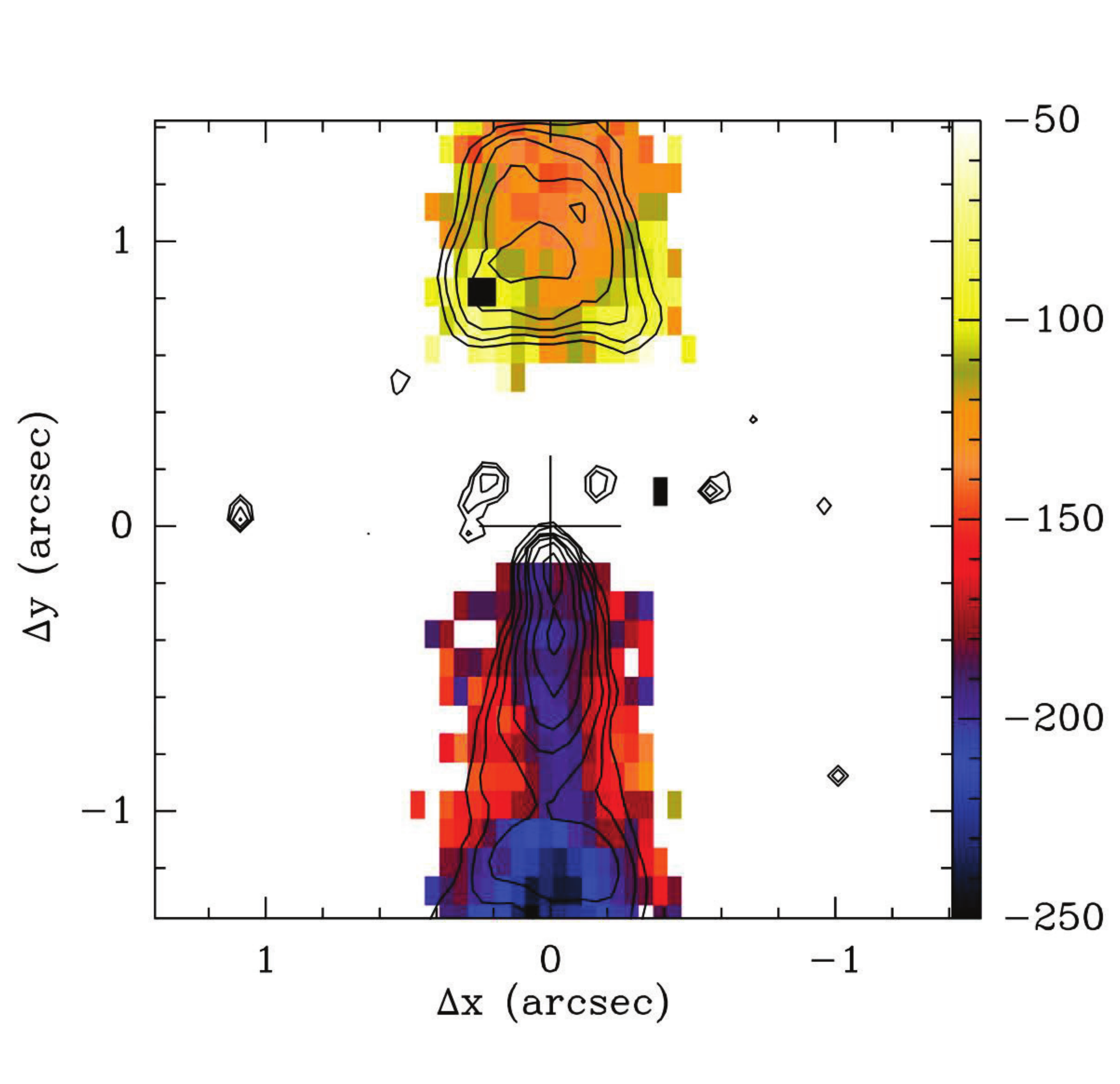}
\caption{\small Map of centroid velocities in the DG Tau jet, as determined from the [Fe II]1.64$\mu$m line observed with the SINFONI IFU on the VLT. Note the fast drop in velocity away from the jet axis, and the velocity asymmetry between blue and red lobes (velocities in the redshifted lobe (on top) were given a minus sign to ease comparison). One arcsecond is 140 pc. From \citet{2011A&A...532A..59A}.} 
\label{fig:velmap}
\end{figure}

Other constraints for jet launching models come from the overall kinematics in the inner few 100 au (where internal shocks and interaction with ambient gas are still limited). Obtaining such information is very demanding as it requires spectro-imaging at sub-arcsecond resolution, either with HST or with powerful Adaptive Optics (AO) systems from the ground, coupled with a long-slit or IFU. It is therefore only available for a handful of bright jets.  A useful result of such studies is that while jet acceleration scales and terminal velocities seem equally compatible with stellar wind, X-wind or D-wind models \citep{2009pjc..book..247C}, there is a  {\em clear drop in velocity towards the jet edges} (as illustrated e.g. in Fig.~\ref{fig:velmap}).  This ``onion-like'' velocity structure, first discovered in the DG Tau jet, has been seen whenever the jet base is resolved laterally and thus may be quite general \citep{2007AJ...133.1221B, 2008ApJ...689.1112C, 2009ApJ...694..654P, 2011A&A...532A..59A}. It argues against the ``classical'' X-wind model where the ejection speed is similar at all angles \citep{2007prpl.conf..261S}, and instead requires that the optically bright jet beam is closely surrounded by a slower wide-angle ``wind". A natural explanation for such transverse velocity decrease is a range of launch radii in an MHD disk wind \citep{2011A&A...532A..59A}, or a  magnetospheric wind surrounded by a disk wind \citep{2009ApJ...694..654P}. Turbulent mixing layers and material ejected sideways from internal working surfaces may also contribute to this low-velocity ``sheath" (see Fig. \ref{Hartigan} and \citet{2008A&A...487.1019G}). Studies combining high spectral and spatial resolution will be essential to shed further light on this issue. 

Asymmetries in jet velocity, density and opening angle between the blue and red lobes are seen in many jets \citep[][see also Fig.~\ref{fig:velmap}]{1994ApJ...427L..99H,2011A&A...527A..13P}. They hold another fundamental clue to the jet launch process, because they remove ill-known variables like stellar mass, disk truncation radius, etc. which are the same for both sides of the flow. Recent studies of RW Aur show that the velocity asymmetry varies over time, while the velocity dispersion remains the same fraction of jet speed in both lobes \citep{2009A&A...506..763M,2009ApJ...705.1388H}. Such asymmetries could be modelled by MHD disk winds where the launch radii or magnetic lever arms differ on either side \citep{2006A&A...453..785F,2010ApJ...714.1733S}.  Possible physical reasons for this are eg. different ionization or magnetic diffusivities on the two faces of the disk \citep{2013ApJ...769...76B, 2013ApJ...774...12F}. Investigating if and how strong blue/red asymmetries can be produced in magnetospheric or stellar winds, along with testable differences compared with D-winds, should be a priority for future theoretical studies.

\subsection{Resolving the Central Engine}
\label{Interferometry}

\begin{figure}[t]
 \epsscale{1.0}
 \plotone{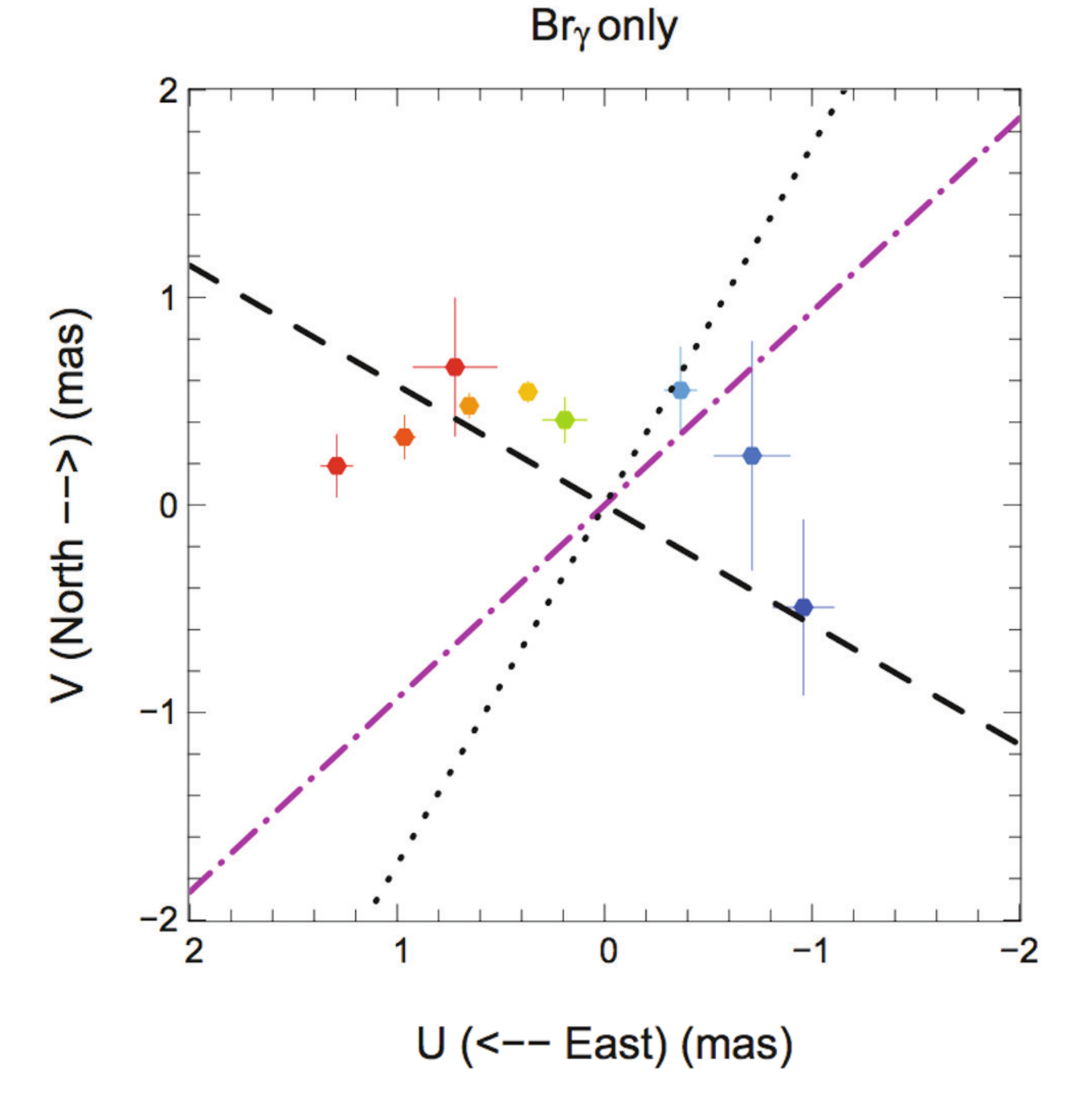}
 \caption{\small The Br\,$\gamma$ 2-D photocenter position in Z CMa as a function of velocity across the Br$\gamma$ profile, using AMBER/VLTI. The velocity goes from -350 (blue) to +350 (red) kms$^{\rm -1}$. The position angle  of the binary (dashed-dotted line), known large-scale outflow (dashed line) and the direction 
 perpendicular to the jet (dotted line) are overplotted. Note that information is being recovered 
 on 0.1 milliarcsecond scales ! From \citet{2010A&A...517L...3B}.}
 \label{Benisty1}
\end{figure}

At the time of PPV, near-infrared (NIR) interferometric measurements in young stars were only possible in the dust continuum, revealing sizes and fluxes compatible with puffed-up rims at the dust sublimation radius 
\citep[see Fig.\,\ref{Hartigan} and][]{2007prpl.conf..539M}.  
MHD disk-winds capable of lifting dust particles have recently been suggested as an alternative means for producing the interferometric sizes and NIR excess in Herbig Ae/Be stars \citep{2012ApJ...758..100B}. The ability to spatially/spectrally resolve Hydrogen lines has also recently been achieved by the VLTI, the Keck Interferometer and the CHARA array.  These studies now enable in-depth studies of 
the spatial distribution and kinematics of the {\em gas} on sub-au scales and bring new constraints on the connection between accretion and ejection. 

Strong Hydrogen emission lines are among the most prominent manifestation of an actively accreting young star. In T Tauri and Herbig Ae stars, they are considered as good proxies 
of mass accretion onto the star, as their luminosity correlates with the accretion rate measured from the UV excess 
(see \citet{2004AJ....128.1294C} and \S\,\ref{Accretion-Outflow}). Yet, their precise origin is still unclear.
Interferometric observations at low spectral resolution (R$\sim$1500-2000) with 
the Keck Interferometer and the AMBER instrument at VLTI provided the first average size measurements in Br${\gamma}$ in about 20 young stars. Gaseous emission is generally more compact than K-band dust continuum (normally located at 0.2-0.5~au).  \citet{2008A&A...489.1157K} fitted typical ring radii $\sim$0.15 to 2.22~au for 5 intermediate-mass young stars. \citet{2010ApJ...718..774E} retrieved smaller extents from 0.04 to 0.28~au for 11 solar-mass and intermediate-mass young stars. 
The common interpretation is that the smallest sizes are dominated by magnetospheric accretion, while 
sizes larger than $\sim$0.1~au trace compact outflows. These results firmly establish the contribution of ejection processes to Hydrogen line formation. 

The connection between accretion and ejection processes on au-scales has recently been specifically addressed in young spectroscopic binaries, where numerical models predict enhanced accretion near periastron. In the close Herbig Ae binary HD104237 (separation $\sim$0.22~au), more than 90\% of the Br$\gamma$ line emission is unresolved and explained by magnetospheric emission that increases at periastron. The large-scale jet should be fed/collimated by the circumbinary disk \citep{2013MNRAS.430.1839G}. The wider, massive Herbig Be binary HD200775 (separation $\sim$5~au) was studied in H${\alpha}$ with the VEGA instrument at the CHARA array. The large size increase near periastron (from 0.2 to $\sim$0.6 au) indicates simultaneously enhanced ejection, in a non-spherical wind \citep{2013A&A...555A.113B}. 
Centroids shifts with 0.1mas precision across the Br$\gamma$ line profile have also been achieved. They reveal a bipolar outflow in the binary Herbig Be star Z~CMa, with a clear connection between its accretion outburst and episodic ejection (see Fig.\,\ref{Benisty1})  \citep{2010A&A...517L...3B}. 
These findings suggest that the accretion-ejection connection seen in T Tauri stars extends well into the Herbig~Ae/Be mass regime. 

Finally, spectrally resolved interferometric observations of the Herbig Ae star AB~Aur in H$\alpha$
\citep{2010A&A...516L...1R} and of the Herbig Be star MWC~297 in Br$\gamma$
\citep{2011A&A...527A.103W} have been modelled using radiative transfer codes 
simulating stellar and/or disk winds. These studies show that HI line emission is 
enhanced towards the equator, lending support to the scenario of magneto-centrifugal launching of jets through disk-winds, rather than through stellar winds. In the near future, spatially resolved multi-wavelength observations of lines emitted at different optical depths (combining e.g., H${\alpha}$ and  Br${\gamma}$) will bring additional constraints. The next generation VLTI imaging instrument GRAVITY is expected to bring the first model-independent images of the central engine in NIR, and to enable statistical studies of solar-mass young stars. 

\subsection{Jets on Small Scales: A High-Energy Perspective}

At the time of PPV, observations of jets and outflows from young stars were 
largely confined to optical or longer wavelength regimes with the occasional foray into the UV. Since then, 
however, jets from protostars and T Tauri stars have been found to contain plasma at 
temperatures of several 
million K. While this discovery came as a surprise, it was not completely unpredictable.
Indeed jet flow velocities of 300-400~km~s$^{-1}$ can, when flowing against a stationary 
obstacle, easily shock-heat gas up to $\approx 1$~MK.  The ensuing X-ray emission would then serve as a valuable jet diagnostic \citep{2002ApJ...576L.149R}. 
However, optical observations of jets within 100 au of their source typically indicate low shock velocities 
$\le 30-100$~km~s$^{-1}$  \citep[e.g.][]{2000A&A...356L..41L,2007ApJ...660..426H}. 
The detection of strong X-rays in jets on small scales
was therefore still not fully anticipated.

\begin{figure}[t]
 \epsscale{1.0}
 \plotone{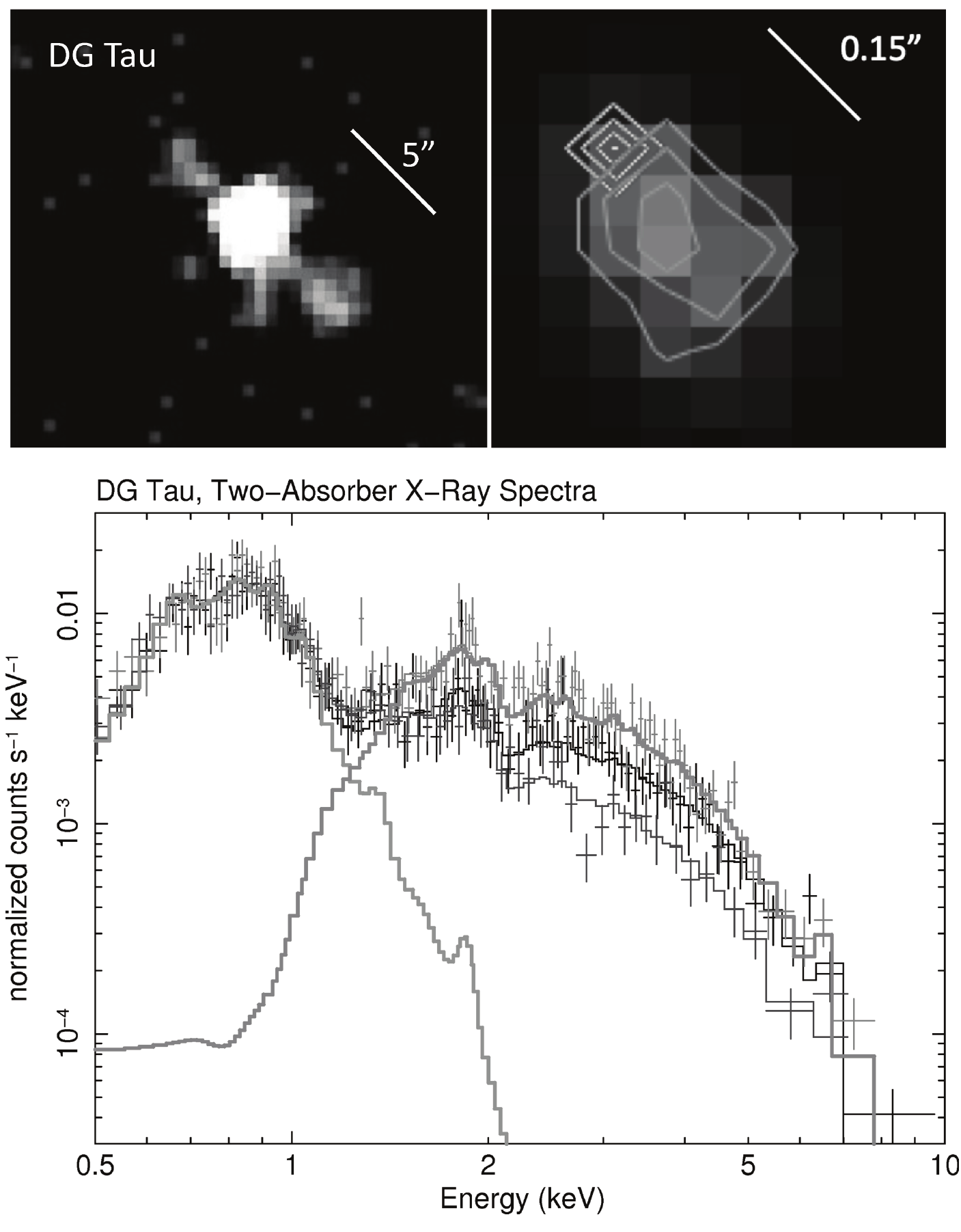}
 \caption{\small The X-ray jet of DG Tau. Upper left: Larger-scale structure
of the jet (SW) and counter-jet (NE) observed by Chandra in
the 0.6-1.7 keV range (G\"udel et al. 2008). 
- Upper right: Innermost region of the soft forward jet (extended gray contours) peaking
30 AU to the SW from DG Tau itself (compact hard source); from G\"udel et al. 2013, in prep. 
- Lower panel: Three Two-Absorber X-ray (TAX) spectra of DG Tau
observed a few days apart. The thick gray histograms show
model fits to the soft (below 1~keV) and the hard (above 1.5~keV)
spectral domains, respectively. The strongly absorbed, variable
hard spectrum originates in the stellar corona; the constant,
little absorbed soft spectrum comes from the jet base in the upper 
right figure (G{\"u}del et al. 2013, in prep.).}
\label{fig:xrays}
\end{figure}

The Taurus jets of L1551 IRS-5, DG Tau, and RY Tau show luminous ($L_{\rm X} 
\approx 10^{28} - 10^{29}$~erg~s$^{-1}$) X-ray sources at distances
corresponding to 30--140~au from the driving star \citep{2002A&A...386..204F,2003ApJ...584..843B,2008A&A...478..797G,2011A&A...530A.123S,2011ApJ...737...19S}. 
A compact X-ray jet was also detected in the eruptive variable Z CMa \citep{2009A&A...499..529S}.
The X-ray spectra of these jet sources are soft, but still require electron temperatures 
of $\approx 3-7$~MK. Further spatially unresolved jets have been discovered based on 
{\it Two-Absorber X-ray} (TAX) spectra. These composite spectra reveal a hard, strongly absorbed spectral component (the star) on top of a soft, little absorbed component (the X-ray jet) \citep{2007A&A...468..515G},
an identification explicitly demonstrated for DG Tau (see Fig.~\ref{fig:xrays}). 
The strongly differing absorption column densities between the two components (by a factor of $\sim$ten)
indicate that the jet X-ray source is located far outside the immediate
stellar environment, where hard coronal X-rays are subject to strong absorption.
So far, objects
like GV Tau, CW Tau, HN Tau, DP Tau and Sz 102 belong to this class \citep{2009pjc..book..347G}; the Beehive proplyd in the 
Orion Nebula cluster is another obvious example \citep{2005ApJS..160..511K}.

The best-studied bright, central X-ray jet of DG Tau has been found to be stationary on timescales
of several years on spatial scales of about 30 au from the central star
\citep[{\em G\"udel et al.}, 2013 in preparation]{2008A&A...488L..13S}. 
Its extent along the jet axis seems to be solely determined by plasma cooling. 
An assessment of the relevant cooling mechanisms \citep{2008A&A...478..797G,2008A&A...488L..13S} suggests that radiative cooling dominates 
for $n_{\rm e} > 10^4-10^5$~cm$^{-3}$, which may be appropriate for these central
sources. For example, an electron density of $n_{\rm e}\approx 10^6$~cm$^{-3}$
and a flow speed of 300~km~s$^{-1}$ are in agreement with the observed 
extension of DG Tau's inner X-ray source of $\approx 0\farcs3-0\farcs5$ as a result
of radiative cooling \citep{2008A&A...488L..13S}. The location of the X-ray source
relative to emission sources at lower temperatures may also be revealing. \citet{2013A&A...550L...1S} obtained high-resolution HST observations in optical and ultraviolet lines and found a high-velocity ($\approx 200$~km~s$^{-1})$ C~{\sc iv}-emitting cloud slightly downwind from the X-ray source, but its luminosity is too high to be explained by cooling plasma previously emitting in X-rays. 
This observation suggests local heating even beyond the X-ray source. 

How is the plasma heated to several MK within tens of au of the central
star? Shocks that produce X-rays require very high jet velocities
 $v_{\rm s} = 370-525$~km~s$^{-1}$ (for $T = 4$~MK) depending on the ionisation
degree of the pre-shock material \citep{2003ApJ...584..843B}. The fixed nature of 
the inner sources suggests they are associated with a stationary heating process in the 
launching or collimation region \citep{2008A&A...478..797G,2008A&A...488L..13S}. 
Viable models include: 
i) Shocks that form when an initially wide wind is deflected and collimated into a jet, 
perhaps by magnetic fields that act as a nozzle for the heated plasma. X-ray luminosity and plasma cooling indicate pre-shock densities of order $10^3 - 10^4$~cm$^{-3}$ \citep{2003ApJ...584..843B}. Specifically, a diamond shock forming at the opening of a (magnetic?) nozzle and 
producing a hot, standing shock was modeled by \citet{2011ApJ...737...54B}.
ii) Randomly accelerated and ejected clouds 
of gas at different velocities produce, through collisions,
chains of moving but also stationary knots along the jet with X-ray emission
characteristics similar to what is observed \citep{2010A&A...517A..68B}. One 
potential drawback of this model is that
very high initial velocities are required to reproduce moderate-velocity 
X-ray knots. 

A major problem with all these shock models is that the high velocities required
to reach the observed temperatures are not observed in any jet spectral lines so far. 
However 
the X-ray emitting plasma component contributes only a minor 
fraction to the total mass loss rate of the associated atomic jet:
$\approx$10$^{\rm -3}$ in DG Tau \citep{2008A&A...488L..13S}. It is 
therefore conceivable that the X-rays are produced in a super-fast but
rather minor jet component not detected at other wavelengths, e.g. a stellar or magnetospheric wind 
\citep{2009A&A...493..579G}. In this context, it may be relevant that X-ray jet
models based on radiative cooling times indicate very small filling factors $f$ of order $10^{-6}$
but high electron densities $n_{\rm e}$, e.g., $n_{\rm e}> 10^5$~cm$^{-3}$ \citep{2008A&A...478..797G};
the resulting pressure would then far exceed that in the cooler $10^4$~K atomic jet, and might
contribute to transverse jet expansion.

It is also conceivable that the standing X-ray structures are not 
actually marking the location of a stationary heating process, but only the exit
points from denser gaseous environments within which X-rays are absorbed and which 
obscure our view to the initial high-energy source. This is an attractive explanation for 
L1551 IRS-5 with its deeply embedded protostellar binary \citep{2003ApJ...584..843B}; 
it could also hold for the soft emission in DG~Tau which is seen to be produced 
near the base of a converging cone of  H$_2$ emitting material that may block the view to the source closer
than 0\farcs15 of the star (\citealt{2013A&A...557A.110S}, {\em G\"udel et al.} 2013, in preparation). With these ideas in mind, an alternative model 
could involve the production of hot plasma in 
the immediate stellar environment through magnetic reconnection of star-disk magnetic 
fields, ejecting high-velocity plasma clouds analogous to solar coronal mass ejections 
\citep{1996ApJ...468L..37H}. If these cool, they may eventually collide with the (slower) jet gas 
and therefore shock-heat gas further out \citep{2011ApJ...737...19S}.

\subsection{Connection with Laboratory Experiments: Magnetic Tower Jets}

A key development since PPV, and of direct relevance to the launching mechanism,
has been the first successful production of 
laboratory jets driven by the pressure gradient of a toroidal magnetic field  
\citep{2005PPCF...47B.465L,2007PhPl...14e6501C}, in a topology similar to the 
``magnetic tower'' model of astrophysical jets 
\citep{1996MNRAS.279..389L,2006MNRAS.369.1167L}. The generated outflow consists of a current-carrying central jet, collimated by strong toroidal fields in a surrounding magnetically-dominated expanding cavity, which in turn is confined by the pressure of the ambient medium. The most recent configurations even allow for the generation of 
several eruptions within one experimental run 
\citep{2009ApJ...691L.147C, 2010PhPl...17k2708S}.  The experiments are scalable 
to astrophysical flows because critical dimensionless numbers such as the plasma 
collisional/radiative cooling parameter ($\chi \simeq 0.1$), and ratio of thermal 
to magnetic pressure ($\beta \simeq 1$), are all in the appropriate ranges. 
Furthermore, the viscous Reynolds number ($Re \simeq 10^6$) and magnetic Reynolds 
number ($Re_M \simeq 200-500$) are much greater than unity, ensuring that transport 
arises predominantly by advection with the flow. 

The main findings from these magnetic tower experiments are the following: 
an efficient conversion of magnetic energy into flow kinetic energy; 
a high degree of jet collimation ($<$10$^{\rm o}$) for sufficiently strong 
radiative cooling; an enhanced collimation for episodic jets, 
as magnetic fields trapped in the previously ejected plasma add to the 
collimation of the later episodes; the generation of an X-ray pulse at 
each new eruption, as the central jet is compressed on-axis by the magnetic 
field; the development of current-driven MHD instabilities leading to variability 
in density ($\simeq$100\%) and velocity ($\simeq$30\%); In particular, these 
experiments show that kink-mode instabilities {\it disrupt but do not destroy the 
MHD jet}, despite a dominant toroidal field. Instead, the non-linear saturation of the unstable modes fragments the beam into chains of dense knots that propagate at a range of velocities around the average beam speed.  Compelling similarities of the episodic jet behaviour in laboratory experiments with observations of  transient bubble-like structures in the XZ~Tau and DG~Tau jets are discussed by \citet{2009ApJ...691L.147C} and \citet{2011A&A...532A..59A}. The stability and possible observational signatures from different configurations of magnetic tower jets were recently studied using numerical simulations with the AstroBEAR code \citep{2012ApJ...757...66H}.  

\begin{figure}[t]
 \epsscale{1.0}
 \plotone{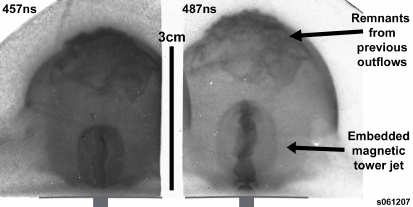}
 \caption{\small Episodic ``magnetic tower'' jet produced in laboratory experiments with the MAGPIE pulsed-power facility. The self-emission XUV images show the growth of the latest magnetic jet and cavity inside the broader cavity created by previous episodes. Adapted from \citet{2009ApJ...691L.147C}.}
\end{figure}

\subsection{Impact on Planet Formation}

About 25\% of the CTTS in Taurus with known jets show detectable X-ray emission from the jet base (G{\"u}del, private communication). This could be relevant for the processing of circumstellar material in their protoplanetary disks. Apart from driving chemical processing, direct heating of the disk surface by X-ray jets may induce photo-evaporation (see Fig.\,\ref{Hartigan}) that competes with that induced by stellar X-rays and UV photons, because of the more favorable illumination geometry. Simple estimates for 
DG~Tau suggest that photo-evaporation outside about 20~au would be dominated by X-rays from the jet  ({\em G\"udel et al.} 2013, in preparation).

At the same time, dusty MHD disk winds if present could effectively screen the disk against the {\em stellar} FUV and Xray photons. For an accretion rate $\simeq 10^{-7} M_\odot$ yr$^{-1}$, an MHD disk wind launched out to 1~au would attenuate stellar photons reaching the disk surface by $A_V \simeq$ 10\,mag and a factor $\simeq$ 500 in coronal Xrays, 
while the star would remain visible to an outside observer with $A_V \leq$ 1\,mag for inclinations
up to $70^o$ from pole-on \citep{2012A&A...538A...2P}. 

Another important dynamical feedback of MHD disk winds on the planet formation zone would be to induce fast radial accretion at sonic speeds due to efficient angular momentum removal by the wind torque (see  the chapter by Turner et al. and references therein), and to modify planet migration through the associated strong magnetic disk fields (see the chapter by Baruteau et al. and references therein).  The thermal processing, coagulation and fall-back of dust grains ejected in an MHD disk wind from 1--3 au was also recently invoked as a means to form and radially redistribute chondrules in our solar system \citep{2012E&PSL.327...61S}.

\section{\textbf{ENVELOPE AND PROPAGATION SCALES (10$^{\rm 2}$au - 0.5pc)}}

\subsection{Jet Physical Conditions Across Star Formation Phases} 
\label{sec:multiple}

Since PPV, the {\em Spitzer}, {\em Herschel} and {\em Chandra} missions along with improved ground-based facilities have allowed to study jets on intermediate scales in younger sources, and in temperature/chemical regimes unexplored in the past. Such studies reveal a frequent coexistence of molecular gas at 500--2000 K with atomic gas at $10^4$ K, and in a few cases with hot plasma at several MK. This broad range of conditions was unanticipated and may arise from several factors: 1) the intrinsic spread of  physical and chemical conditions present in the cooling zones behind radiative shocks, 2) the interaction of the jet with its environment (eg. entrainment of molecules along the jet beam), 3) the simultaneous contributions of (possibly molecular) disk winds, stellar winds, and magnetospheric ejections. Disentangling these 3 factors is essential to obtain accurate jet properties and to understand its interaction with the natal core. The evolution of jet composition as the source evolves from Class 0 to Class II holds important clues to this issue. 

Molecular jets from the youngest protostars (class 0 sources) have benefited most from recent progress in the sub/mm and IR ranges. Interferometric maps in CO and SiO show that they reach de-projected velocities of several hundred km/s, 
as expected for the ``primary wind'' from the central source \citep{2007prpl.conf..245A, 2007A&A...462L..53C, 2010ApJ...717...58H}. This is further supported by chemical abundances that are clearly distinct from those in swept-up ambient gas \citep{2010A&A...522A..91T}.  Multi-line SiO observations show that they are warm (T$_{kin}$ in the range 100-500 K) and dense (n(H$_2$) $\ge$ 10$^5$--10$^6$ cm$^{-3}$) \citep{2007A&A...462..163N, 2007A&A...468L..29C}. This result was confirmed via H$_2$ mid-IR observations of the class 0 jets L1448 and HH211 \citep{2009ApJ...692....1D, 2010A&A...521A...7D} and {\em Herschel} observations of water lines in L1448 \citep {2011A&A...531L...1K, 2013A&A...549A..16N}.  ALMA observations will soon provide an unprecedented view of these warm, dense molecular jets, as already illustrated by first results in the CO (6-5) line \citep[e.g. ][]{2013A&A...549L...6K, 2013MNRAS.430L..10L}. In particular they should clarify the corresponding ejection/accretion ratio \citep[currently subject to significant uncertainties, eg.][]{2010ApJ...713..731L}.

\begin{figure}[t]
\epsscale{1.0}
\plotone{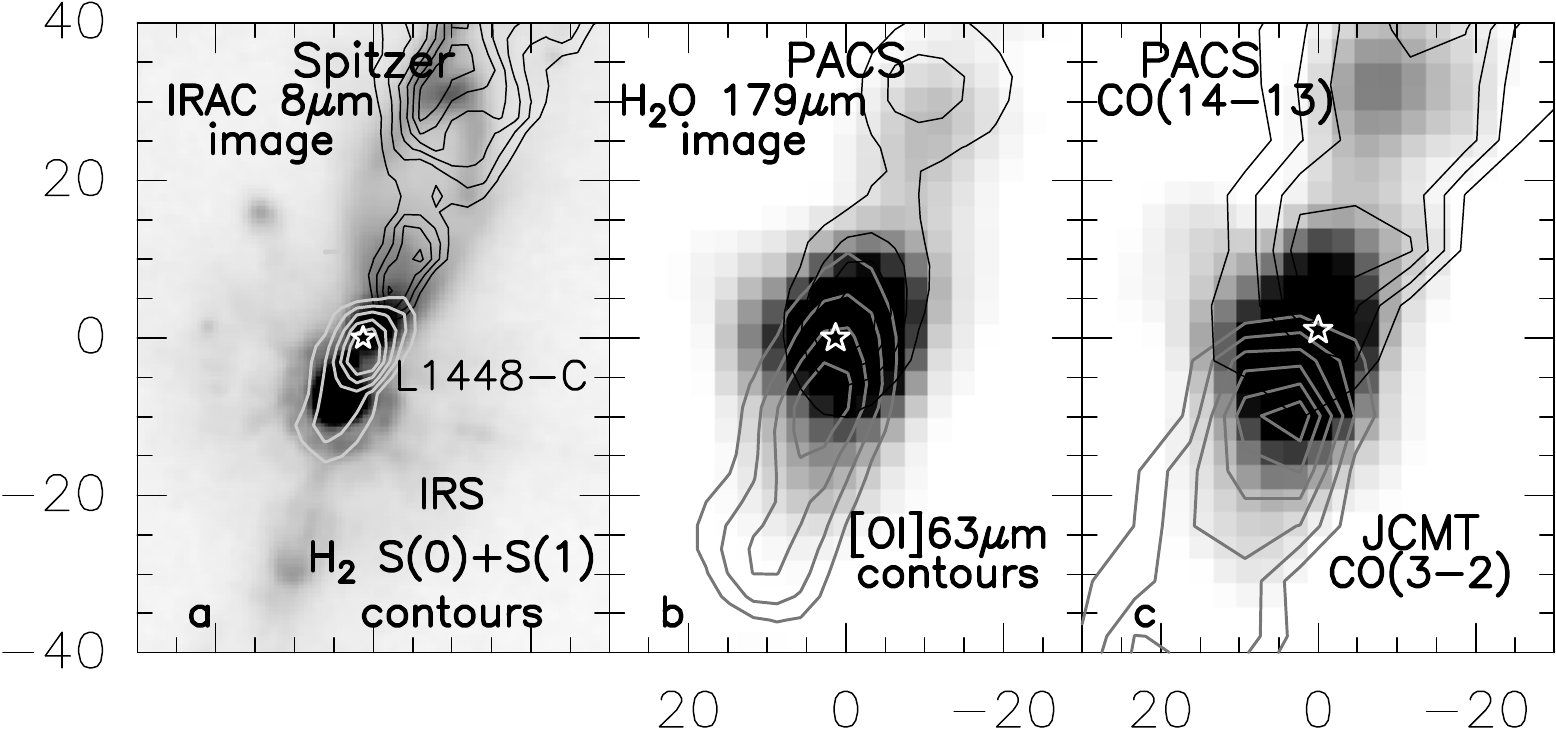}
\caption{\small {\em Spitzer} and {\em Herschel} spectral images of the jet from the class 0 object L1448-C, revealing the spatial distribution of warm molecular gas at 300-2000 K and the presence of an embedded bipolar atomic jet: a) contours of H$_2$ S(0) and S(1) line emission from {\em Spitzer} IRS superimposed on the IRAC 8$\mu$m  image in greyscale; b) blue-shifted (black contours) and red-shifted (gray contours) emission of [OI] 63$\mu$m superimposed on a greyscale image of the H$_2$O 179$\mu$m line, both obtained with {\em Herschel}/PACS; c) greyscale PACS image of the CO(14-13) emission with superposed contours of CO(3-2) from JCMT. Panel a) is adapted from \citet{2011ApJ...738...80G}. Panels b and c from \citet{2013A&A...549A..16N} and Nisini et al. 2013 (in prep.) }
\label{fig:nisini}
\end{figure}

{\em Spitzer} also revealed for the first time an embedded {\em atomic} component associated with these molecular Class 0 jets, via mid-IR lines of [Fe II], [S I] and [Si II] \citep{2009ApJ...692....1D, 2010A&A...521A...7D}. It is characterized by a low electron density ($\sim$ 100-400 cm$^{-3}$), moderate ionization fractions of $<$10$^{-3}$ and T$<$3000 K. However, its contribution to the overall jet dynamics and its relationship to the molecular jet are still very uncertain.  This issue will be likely revised thanks to {\em Herschel} PACS observations that resolve strong collimated [OI] 63$\mu$m emission in several Class 0 jets (see Fig.~\ref{fig:nisini}), this line being a better tracer of mass-flux. 

On larger scales, the shocks caused by the interaction of Class 0 jets with the ambient medium have been probed with much better resolution and sensitivity than possible in the 1990s with the ISO satellite. 
{\it Spitzer} line maps demonstrate a smooth H$_2$ temperature stratification between 100~K and 4000~K \citep{2009ApJ...706..170N} where H$_2$ pure rotational lines are the main cooling agent \citep{2010ApJ...724...69N}. The other two main molecular shock coolants, CO and H$_2$O, were studied in detail with {\em Herschel}. The H$_2$O and  far-IR CO lines (at $J \ge 13$) are strictly correlated with H$_2$ $v=0$ and trace high pressure post-shock gas \citep{2012A&A...538A..45S, 2013A&A...551A.116T}. Contrary to simple expectations, the contribution of water to total cooling is never larger than $\sim$ 20-30\% \citep{2010A&A...518L.120N} (see chapter by van Dishoek et al. for a more general discussion of water). Detailed analysis of CO and H$_2$O line profiles and maps reveals multiple shock components within 30\arcsec, with different temperatures, sizes, and water abundances that further complicate the analysis \citep[e.g.][]{2012ApJ...757L..25L,2013A&A...557A..22S}. As to more complex organic molecules, their abundances and deuteration levels in outflow shocks have proven to offer a useful ``fossil" record of ice mantles formed in the cold preshock ambient cloud \citep{2008ApJ...681L..21A,2012ApJ...757L...9C}.


The dissociative ``reverse shock'' (or ``Mach disk'') where the jet currently impacts on the leading bowshock (see Fig. \ref{Hartigan}) was also unambiguously identified for the first time in Class 0 jets, in [Ne~II], [Si~II] and [Fe~II] lines with {\em Spitzer} \citep{2006ApJ...649..816N, 2008ApJ...680L.117T, 2012ApJ...751....9T}, and in OH and [O~I] with {\em Herschel}/PACS \citep{2012A&A...539L...3B}. In the latter case, the momentum flux in the reverse shock seems sufficient to drive the whole swept-up CO cavity. 

The abundance, excitation, and collimation of molecules in jets clearly evolve in time. In contrast to Class 0 sources,  older Class I jets are undetectable (or barely so) in low-J CO and SiO emission. Hot molecular gas at $\simeq$1000-2000 K is still seen, in the form of ro-vibrational H$_2$ emission and more rarely $v = 0-1$ CO absorption \citep[see][and refs. therein]{2011A&A...528A...3D,2011A&A...533A.112H}.  While some $H_2$ may be associated with the fast atomic jet,  it mainly traces a slower ``intermediate velocity component'' (IVC) $\simeq 10-50$ km/s near the jet base, and in all cases carries a 10-1000 times smaller mass flux than the atomic jet \citep[][]{2005A&A...441..159N, 2006A&A...456..189P, 2008A&A...487.1019G, 2011A&A...528A...3D}. In the later Class II stage, hot H$_2$ generally peaks at even smaller velocities $\le 15$ km/s and traces a wider
flow around the atomic jet \citep[see e.g.][]{2006ApJS..165..256H, 2007ApJ...670L..33T,2008ApJ...676..472B, 2013A&A...557A.110S}. 

Concerning the atomic jet component in Class I jets, optical and near-IR line ratios indicate similar temperatures $\simeq 10^4$ K and ionisation fraction $x_e \sim$ 0.05-0.9 as in Class II jets, indicating moderate shock speeds $\sim$ 30-70 km/s, but with higher electron and total density \citep{2005A&A...441..159N, 2006A&A...456..189P,2008A&A...479..503A, 2008A&A...487.1019G}. 
This implies a higher mass flux rate (although the ejection to accretion ratio remains similar, see \S \ref{Accretion-Outflow}).
A more complete view of the different excitation components present in jet beams can be obtained by combining emission lines in a wide wavelength range of 0.3--2$\mu$m. The first such studies in Class II jets reveal a broader range of ionization states (including [S III] and [O III]) than seen in optical lines, probing faster shocks $\ge $100 km/s which must be accounted for in mass-flux rate derivations \citep{2011ApJ...737L..26B}. 

Finally, extended X-ray emission has been resolved with {\em Chandra} along the L1551-IRS5 (Class I) and DG Tau (Class II) jets out to distances of 1000~au, revealing hot plasma at several MK that was totally unanticipated from optical data on similar scales. 
X-rays have been detected as far as 0.1pc to 2.5pc from the driving source, associated with high-excitation Herbig-Haro (HH) objects.  The relationship between X-ray and optical emission, however, is not always clear. HH 80/81 shows X-rays, radio continuum, and optical lines all coinciding at arcsecond resolution. The X-rays, however, point to a factor of 10 lower density and  2--5 times lower speed than the jet  \citep{2004ApJ...605..259P}. An inverse situation is encountered in the Cepheus A East/West region \citep{2005ApJ...626..272P}. Here, the required shock speeds are comparable to flow speeds, but the head of the expanding region is detected in H$\alpha$ and not in X-rays. Hence the hot plasma appears to be heated significantly upstream of the leading working surface, possibly in a reverse shock \citep{2009A&A...508..717S} or in a collision with another jet \citep{2009ApJ...692..943C}. Clearly, further work is needed to fully understand the link between optical and Xray emission from jets and HH objects on intermediate scales. 

\subsection {Magnetic and chemical diagnostics on intermediate scales} 
\label{sec:chemistry} 

Modern models of jet launching all invoke magnetic fields to achieve the desired terminal velocities and narrow collimation angles of $\sim$ 5 degrees. However, measuring field strengths within bright optical jets has proved very difficult because Zeeman splitting is undetectable in optical lines. \citet{2010Sci...330.1209C} were able to measure polarized synchrotron radio emission in the shocked jet of  HH 80/81 and inferred an average field of $\sim$ 200$\mu$G at 0.5~pc, with a helical structure about the jet axis. But this jet, driven by a massive protostar, is quite exceptional by its speed (1000 km/s) and brightness in the radio range. 

In optical jets, it is still  possible to estimate B-fields  through the effect they have on post-shock compression and the resulting emission line ratios. \citet{1992ApJ...399..231M, 1993ApJ...410..764M} inferred a preshock field of $\sim 20-30\mu$G in distant bowshocks of two Class I jets with a density of 100--200~cm$^{-3}$. More recently, \citet{2009A&A...507..581T, 2012ApJ...746...96T} estimated B $\simeq$ 500$\mu$G at $n_H \simeq 1-5 \times 10^4$ cm$^{-3}$ in two Class II microjets within 500 au of the source. These all yield transverse Alfv\' en speeds $V_{\rm A,\phi} \simeq$ 4 km/s, typically 1/50th of the jet speed. This value is a lower limit, as the B-field could have been partly dissipated by reconnection or ambipolar diffusion, between the point where the jet is launched to where shock waves are observed, and further lowered by ``velocity stretching'' between internal working surfaces \citep{2007ApJ...661..910H}. Hence these results provide interesting constraints for MHD jet launching models. Resolved spatial maps of the ionization fraction, temperature, and density obtained with HST for the HH 30 jet in two epochs \citep{2007ApJ...660..426H} also reveal an unexplained new phenomenon where the highest ionization lies upstream from the emission knots and does not show a correlated density increase (such behavior was also observed by HST in RW Aur \citep{2009A&A...506..763M}). Models of line emission from magnetized jet shocks have yet to fully confront these observational constraints.

The magnetic field in {\em molecular} jets from Class 0 sources could also, in principle, 
be constrained by shock modeling. Such efforts are complicated by the fact that two kinds of shocks may exist: the sudden ``J-type'' shock fronts and the broader C-type shocks where the magnetic field is strong enough to decouple ions and neutrals and energy is dissipated by ambipolar diffusion. Given uncertainties in beam filling factor, H$_2$ data alone are often insufficient  to constrain the B value, unless it is large enough to make C-shock cooling regions spatially resolvable \citep[e.g. as in Orion BN-KL][]{2008A&A...477..203K, 2010A&A...513A...5G} or when the transition from C to J-type shock can be located along a large bowshock surface \citep[e.g.][]{2008A&A...481..123G}.  However, B values in bowshocks may be more relevant to the external medium than to the jet itself. Shock chemistry  offers additional clues \citep[see the excellent review of this topic in][]{2007prpl.conf..245A} but requires complex modeling. For example, SiO was long believed to offer an unambiguous tracer of C-shocks, but recent models now also predict substantial SiO in dense J-shocks, from grain-grain shattering \citep{Guillet2009}. One must also account for the fact that young C-shocks in jets will contain an embedded J-type front. {\it Herschel} observations bring additional C/J-shock diagnostics such as [O I], OH, and NH$_3$ lines \citep{Flower2013} so that our understanding of B-field in class 0 jets should greatly progress in the coming years. 


Another indirect clue to the launch region is whether the jet is depleted in refractory elements, as one would expect if it originates from beyond the dust sublimation radius ($\simeq 0.1-1$ au) where these elements would be mainly locked up in grains. Refractory gas-phase abundances have been measured in bright HH objects for decades \citep[e.g. ][]{1981ApJS...47..117B}. Only recently however have such studies been extended to the much fainter jet beams. Assuming solar abundances, measurements indicate significant gas-phase depletions of Fe, Si and/or Ca   \citep{2005A&A...441..159N, 2006A&A...456..189P, 2009A&A...506..779P, 2011A&A...527A..13P, 2009ApJ...692....1D, 2010A&A...521A...7D, 2011A&A...532A..59A}. While the data are not extensive and the measurements difficult, the general consensus indicates dust exists in {\em atomic} jets at all evolutionary stages (from Class 0 to Class II), in larger amounts at lower velocities, and gets progressively destroyed along the jet in strong shocks. The dust could be entrained from the surrounding cloud, or may be carried along with gas ejected from the circumstellar disk (see Fig. \ref{Hartigan}). In any case, a large fraction of the dust should survive the acceleration process. Such measurements also argue against the lower velocity gas tracing sideways ejections from internal jet working surfaces; if this were the case, it should be more shock-processed and less depleted in refractories than high-velocity gas, whereas the opposite is observed \citep{2011A&A...532A..59A}. 

The dust content is more difficult to constrain in the {\it molecular} component of Class 0 jets. SiO is the only detected molecule involving a refractory species, and unfortunately it is optically thick in the inner 500 au's of Class 0 jets \citep{2007A&A...468L..29C}. The lower limit on SiO gas-phase abundance is $\simeq$ 10\% of elemental silicon, still compatible with an initially dusty jet  \citep{2012A&A...548L...2C}.  One possible indirect indication that molecular jets might arise from {\em dusty} MHD disk winds are the predicted chemical and temperature structures \citep{2012A&A...538A...2P}. When ionisation by coronal Xrays is included, ion-neutral coupling is sufficient to lift molecules from the disk without destroying them, while efficient dust shielding enables high abundances of H$_2$, CO and H$_2$O. As the wind density drops in the Class I and II phases, dust-shielding is less efficient.  The molecular region moves to larger launch radii $\simeq 0.5-1$ au, while heating by ion-neutral drag increases. This trend would agree with observations of decreasing speed, mass-flux, and collimation of H$_2$ and increasing temperatures in Class 0 to Class II jets (see Sect.~\ref{sec:multiple}). The broad H$_2$O line wings recently discovered towards Class 0 and Class I sources with {\it Herschel}/HIFI \citep{2012A&A...542A...8K} can also be reproduced by this model as well as the correlation with envelope density ({\em Yvart et al. 2013}, in prep.). ALMA and infrared IFUs with laser guide stars will bring key constraints on this scenario and on the origin of molecular jets, in particular through more detailed characterization of their peculiar chemical abundances \citep[c.f.][]{2010A&A...522A..91T} and the confrontation with model predictions.


\subsection {Ejection variability and implications for source and disk properties}
\label{sec:precess} 

Since jets are accretion-driven, outflow properties that change with distance from the source provide important constraints on past temporal variations in the ejection/accretion system, over a huge range of timescales from 
$<5$  to $10^5$ yrs that cannot be probed by any other means.  

Optical, infrared, and millimeter observations show that both atomic and molecular jet beams exhibit a series of closely spaced inner ``knots'' within 0.1pc of the source (see Fig \ref{Hartigan}), together with more distant, well separated larger bows or ``bullets'' which in most cases have a clear correspondence in the opposite lobe, HH212 being the most spectacular example to date \citep{1998Natur.394..862Z}. Atomic jet knots and bows have line ratios characteristic of internal shock waves. Therefore, they cannot just trace episodes of enhanced jet density, which alone {\em would not produce shocks}. Significant variations in speed or ejection angles are also required. Several lines of evidence imply that these shocks are caused by supersonic velocity jumps where fast material catches up with slower ejecta \citep[e.g. ][]{2002ApJ...565L..29R, 2005AJ....130.2197H}. The same was recently demonstrated for their CO ``bullet'' counterparts \citep{2009A&A...495..169S, 2010ApJ...717...58H}. 

The most natural origin for such velocity jumps is initial variability in the ejection speed \citep{1990ApJ...364..601R}. This is supported eg. by numerical simulations of the resulting jet structure, and by HST proper motions at the base of the HH34 jet clearly showing a velocity increase of 50 km/s over the last 400 yrs \citep{2012ApJ...744L..12R}.  The knot/bow spacing and velocity patterns in Class I atomic jets then suggest that up to 3 modes of velocity variability are present in parallel, with typical periods of a few 10, a few 100, and a few 1000 yrs respectively and velocity amplitudes of 20-140 km/s \citep{2002A&A...395..647R}. A strikingly similar hierarchy of knot/bullet spacings is seen in Class 0 jets, suggesting a similar variability behavior \citep[see e.g.][]{2002EAS.....3..147C}. Time-series of Taurus Class II jets at 0\farcs15 resolution show new knots that emerge from within 50-100 au of the source with an even shorter interval of 2.5-5 yrs \citep{2007ApJ...660..426H, 2011A&A...532A..59A}. {\it Spitzer} observations further reveal that the 27yr period knots in HH34 are synchronized to within 5 years between the two jet lobes, implying that the initial perturbation is less than 3 au across at the jet base \citep {2011ApJ...730L..17R}. 

These results set interesting constraints for jet launching and variable accretion models. Proposed physical origins for quasi-periodic jet variability include: stellar magnetic cycles or global magnetospheric relaxations of the star-disk system (3-30yrs), perturbations by unresolved (possibly excentric) binary companions, and EXOr-FUOr outbursts (see chapter by Audard et al.).  Dedicated monitoring of at least a few prototypical sources should be a priority to clarify the link between these phenomena and jet variability.   We note that care must be taken in interpreting such result, however, as jet are likely to be inherently clumpy on sub-radial scale.  The internal dynamics of clumps of different size and velocity represents an essentially different form of dynamics than pure velocity pulsing across the jet cross-section \citep{2012ApJ...746..133Y}. In particular as clumps collide and potentially merge they can mimic the appearence of periodic pulsing \citep{2009ApJ...695..999Y}.

\begin{figure*}[t]
\epsscale{1.7}
\plotone{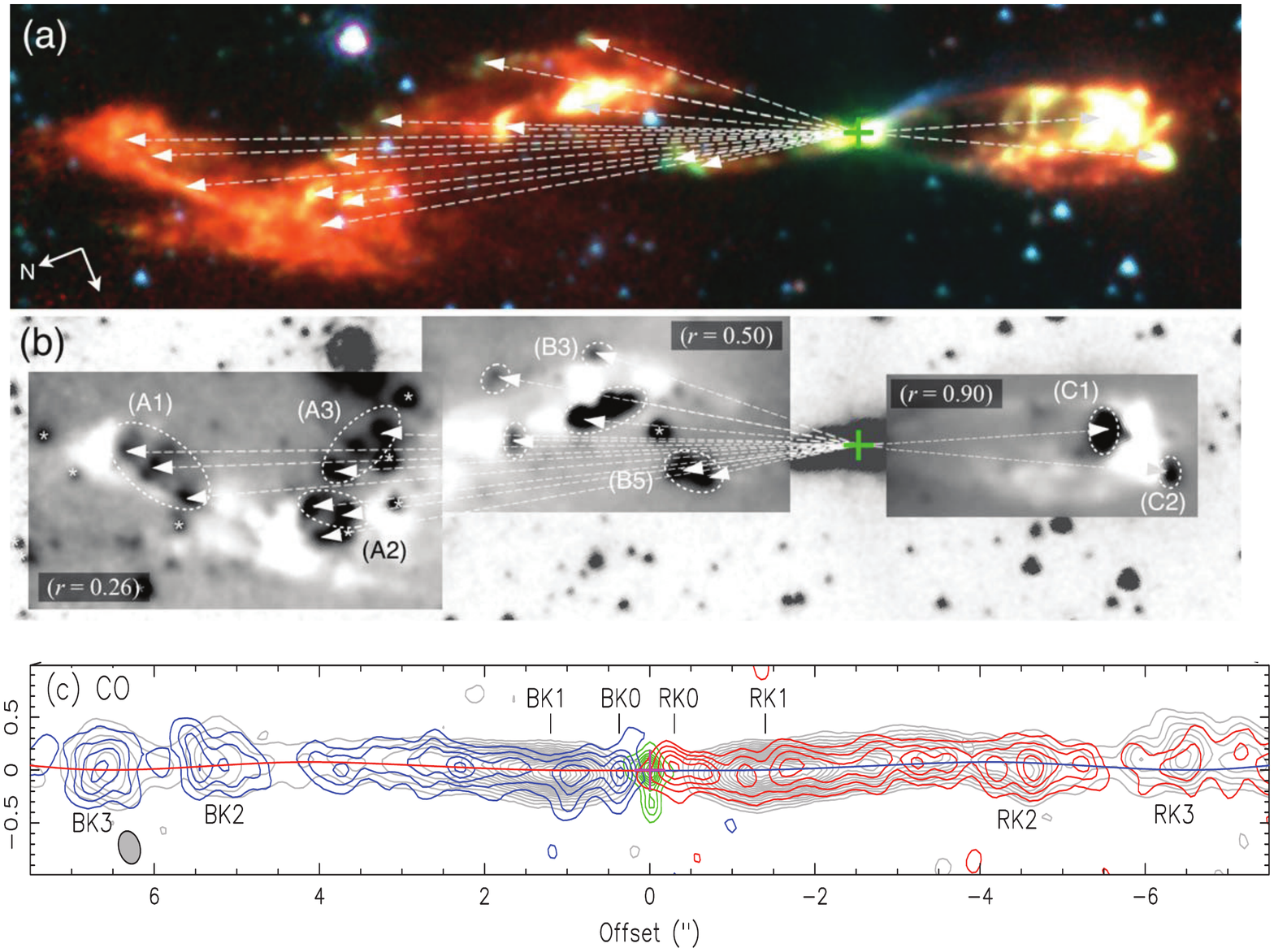}
\caption{\small Examples of S-shaped (precession) and mirror-symmetric (orbital) wiggling in protostellar jets : (a) Three-color {\em Spitzer} IRAC image of the L1157 outflow (blue, green, and red for 3.6, 4.5, and 8.0 $\mu$m, respectively); (b) Difference image of L1157 where warm/dense H$_2$ knots show up in black and are connected by arrows to the central protostar (green cross). (c) jet orbital motion model (curve) superposed onto the CO and SiO knots in the HH211 jet (contours). Adapted from \citet{2011ApJ...743..193T} and \citet{2010ApJ...713..731L}.} 
\label{fig:precess}
\end{figure*}

Internal shocks may also be produced without velocity pulsing if the jet axis wanders sufficiently that dense packets of gas can shock against ambient or slow cocoon material \citep[e.g. ][]{2001MNRAS.322..166L}. Jet axis wandering with time is indeed a common characteristic among sources of various masses and evolutionary stages. Mass ejected from a young stellar object should follow, approximately, a linear trajectory once it leaves the star-disk system, unless it is deflected by a dense clump or a side-wind. And indeed, most knot proper motions are radial to within the errors \citep[e.g. ][]{2005AJ....130.2197H}. Hence, jet wiggles or misaligned sections, commonly seen in the optical, IR and millimeter, most likely indicate a variation in ejection angle. Jet precession produces point-symmetric (S-shaped) wiggles between the jet and counterjet, while orbital motion of the jet source in a binary system will produce mirror-symmetric wiggles. Precession by a few degrees has long been known in Class 0/I jets (see e.g. Fig.~\ref{fig:precess}a-b), with typical periods ranging from 400 to 50,000 yrs \citep[e.g.][]{1996AJ....112.2086E, 1996A&A...307..891G, 1997AJ....114.2095D}, and larger axis changes of up to 45$^o$ in a few sources \citep[e.g.][]{2009ApJ...692..943C}. Mirror-symmetric signatures of jet orbital motion have been identified more recently eg. in the HH211 Class 0 jet (see Fig.~\ref{fig:precess}c), the HH111 Class I jet, and the HH30 Class II jet, with orbital periods of 43 yrs, 1800 yrs, and 114 yrs respectively \citep{2010ApJ...713..731L, 2011ApJ...732L..16N, 2012AJ....144...61E}. 
It is noteworthy that secular disk precession driven by tidal interaction with the orbiting companion (assumed non coplanar) could explain the longer precession timescale observed on larger scales in HH111 \citep{1999ApJ...512L.131T, 2011ApJ...732L..16N}.  Such a coincidence suggests that jet axis precession is due to precession of the disk axis, rather than of the stellar spin axis. Although more examples are needed to confirm this hypothesis, it supports independent conclusions that jet collimation (and possibly ejection) is controled by {\em the disk B-field} (see Section~\ref{Collimation}). Observations of jet orbital motions also provide unique constraints on the mass and separation of close companions which would be otherwise difficult to resolve. Interestingly, the inferred binary separation of 18 au in HH30 is consistent with the size of its inner disk hole \citep{2012AJ....144...61E}. The jet from the Herbig Be member of ZCMa shows wiggles with a 4-8 yr period similar to the timescale of its EXOr outbursts, suggesting that such outbursts may be driven by a yet undetected companion \citep{2010ApJ...720L.119W}. 

\subsection{Jet propagation and shock structure: connection with laboratory experiments}
\label{sec:propagation}

Another major development in the time-domain has been the acquisition of multiple-epoch emission-line images from HST, which now span enough time (10 years) to reveal not only proper motions of individual knots, but also to begin to show how the shock waves evolve and interact. Images of the classic large-scale bowshocks in HH1\&2, HH~34, and HH~47 \citep{2011ApJ...736...29H} show evidence for a variety of phenomena related to jet propagation (see Fig \ref{Hartigan}), including standing deflection shocks where the precessing jet beam encounters the edges of a cavity, and where a strong bow shock encounters a dense obstacle on one side. Knots along the jet may brighten suddenly as denser material flows into the shock front, and fade on the cooling timescale of decades. Multiple bow shocks along working surfaces sometimes overlap to generate bright spots where they intersect, and the morphologies of Mach disks range from well-defined almost planar shocks to small reverse bow shocks as the jet wraps around a denser clump. The bow shocks themselves exhibit strong shear on the side where they encounter slower material, and show evidence for Kelvin-Helmholtz instabilities along the wings of the bows. 

In support to the interpretation of observations, innovative laboratory experiments on jet propagation have been carried out, where magnetic fields are not present or not dynamically significant.  Experiments on pulsed-power facilities investigated the gradual deflection (bending) of supersonic jets by the ram pressure of a side-wind \citep{2004ApJ...616..988L,2005PPCF...47B.465L}. The experimental results were used to benchmark numerical simulations and the same computer code was used to simulate astrophysical systems with scaled-up initial conditions  \citep{2008ApJ...678..968C}. Both the experiments and the astrophysical simulations show that the jet can be deflected by a significant angle ($\simeq 30^\odot$) without being destroyed. The interaction between the jet and the side-wind also leads to variability in the initially laminar flow, driven by the onset of Kelvin-Helmholtz instabilities.

Experiments with laser-driven jets \citep{2010AAS...21620502F,2011PhPl...18h2702H} have been primarily devoted to studies of hydrodynamical instabilities in a jet interacting with, and deflected by, localised dense obstacles \citep{2009ApJ...705.1073H}, in a geometry similar to the HH110 jet. The experimental results have been compared in detail with numerical simulations and show good agreement. Another recent experiment \citep{2012ApJ...746..133Y} investigated the formation of Mach stems in collisions between bow-shocks, which is relevant to observations of similar structures in HH objects with HST. Such collisions are expected in the interaction of clumpy jets with ambient gas, and could lead to the formation of shocks normal to the flow and a localised increase in emission \citep{2011ApJ...736...29H}.

Finally, several experiments \citep{2008PhPl...15h2701N,2012PhPl...19b2708S} investigated the jet-ambient interaction in conditions where radiative cooling is very strong ($\chi \ll 1$). It was found that small-scale clumps, attributed to cooling instabilities, rapidly develop in the bow-shock region and that the clump size decreases for increasing radiative cooling \citep{2013HEDP....9..141S}. These experiments represent the first investigation of cooling instabilities evolving into their highly non-linear stages, which may have observable consequences e.g. on line ratios of high vs. low ionization stages. 

\subsection{Core-to-star efficiency and envelope dissipation}
\label{sec:cavity}

A comparison of the prestellar core mass function with the initial mass function suggests that only 1/3 of the core mass ends up into the star (see eg. chapters by Offner et al. and Padoan et al. in this volume). Since circumstellar disks are seen to contain only a small fraction of the final stellar mass, protostellar jets / winds are prime candidates to explain this low ``core-to-star'' efficiency \citep[eg.][]{2008ApJ...687..340M}. 

An attractive possibility is that a substantial fraction of infalling core gas is re-ejected during early Class 0 collapse via magnetically driven outflows. 3-D MHD simulations of rotating collapse over $3 \times 10^4$ yrs suggest that MHD ejection results in a final accreted mass of only $\simeq 20\%/cos\alpha$ of the initial core mass, where $\alpha$ is the initial angle between the B-field and the core rotation axis \citep{2010MNRAS.409L..39C}. Longer simulations extending to $\simeq 10^5$ yrs with $\alpha = 0$ suggest that mass accretion during the Class I phase brings the final core-to-star efficiency closer to 50\% \citep{2013MNRAS.431.1719M}. Although slightly larger than the observed 30\%, this result indicates that early protostellar MHD ejections could play a key role in determining the core-to-star efficiencies. The ejected mass in this early phase is at relatively low velocity and may constitute part of the low-velocity V-shaped cavities later observed around Class 0 jets (see chapter by Li et al.). Slow outflows recently attributed to very young first or second hydrostatic cores should provide a test of this scenario. 

Another complementary scenario is that swept-up outflow cavities driven by wide angle winds halt infall  by dispersing the infalling envelope \citep{2008ApJ...687..340M}; Early evidence suggestive of envelope dispersion by outflows was presented in PPV \citep{2007prpl.conf..245A}. Here we discuss new elements relevant to this issue, and their resulting implications. Interferometric CO observations in a sample of nearby protostars have demonstrated an increasing opening angle of the outflow cavity with age \citep{2006ApJ...646.1070A}. Class 0 cavities show opening angles of $20^o-50^o$, Class I outflows show $80^o-120^o$, and Class II outflows show cavities of about 100$^o$ to 160$^o$. 
A similar trend is seen in scattered light \citep{2008ApJ...675..427S}. These results may be understood if protostellar winds are wide-angled with a denser inner part along the outflow axis. At early times, only the fastest and densest axial component of the wind punctures the circumstellar environment. As the protostar evolves, entrainment by the outflow decreases the density in the envelope, allowing material at larger angles from the outflow axis to be swept up and widening the outflow cavity \citep{2007prpl.conf..245A}. 3D simulations and synthetic CO observations of protostellar outflows in a turbulent core do show a gradual increase of the cavity opening angle, up to $50^o$ at $5 \times 10^{4}$ yrs, even though the angular distribution of injected momentum remains constant over time \citep{2011ApJ...743...91O}. 

A caveat to this interpretation comes from recent studies claiming mass-flux rates in CO outflows are too low to disperse, alone, the surrounding envelopes within their disappearance timescale of $\simeq 2-3 \times 10^5$ yrs \citep{2007A&A...472..187H, 2010MNRAS.408.1516C}. If this were indeed the case, the observed broadening of outflow cavities with age would then be a {\em consequence, rather than the cause,} of envelope dispersal. Efforts are needed to reduce uncertainties in line opacity, gas temperature, and hidden gas at low-velocity or in atomic form \citep[e.g.,][]{2007A&A...471..873D} to obtain accurate estimates of outflow rates on the relevant scales. It is also noteworthy that the envelope mass typically drops by an order of magnitude between the Class 0 and Class I phases \citep{1996A&A...311..858B}, while the cavity full opening angle $\theta \le 100^o$ encompasses a fraction $1-\cos(\theta/2) \le $ 36\% of the total envelope solid angle, and an even smaller fraction of the envelope {\em mass} (concentrated near the equator by rotational and magnetic flattening).  This seems to indicate that outflow cavities will be too narrow during the early phase where stellar mass is assembled to affect the core-to-star efficiency, although they could still be essential for dissipating the residual envelope at later stages.

Another open question that bears more on the issue of the launching mechanism is the nature of the wide-angle component responsible for the observed opening of outflow cavities. While a  {\em fast} $\simeq 100$ km/s wide-angle wind has often been invoked \citep{2006ApJ...649..845S, 2007prpl.conf..245A}, this now appears ruled out by recent observations indicating a fast drop in velocity away from the jet axis (see Section~\ref{sec:chemistry}). On the other hand, a {\em slow} wide-angle wind may still be present. In particular, an MHD disk wind launched out to several au naturally produces a slow wide-angle flow around a much faster and denser axial jet \citep[see e.g.][]{2007prpl.conf..277P,2012A&A...538A...2P}, with an angular distribution of momentum similar to that used in \citet{2011ApJ...743...91O}. Sideways splashing by major working surfaces (see Section~\ref{sec:precess}) could also contribute to gradually broaden the cavity base.  {\em Herschel} studies of warm $> 300$ K molecular gas have started to reveal the {\em current} shock interaction between the jet/wind and the envelope \citep{2013A&A...557A..23K}.  ALMA maps of outflow cavities promise to shed new light on this issue, thanks to their superb dynamic range and sensitivity to faint features \citep{2013ApJ...774...39A}.


\section{\textbf{PARENT CLOUD SCALES (0.5 - 10$^{\rm 2}$\,pc)}}

\subsection{ Parsec-scale Jets, Outflows and Large Scale Shells}
\label{Large-Scale}

Optical and near-infrared wide-field camera surveys in the late 1990's and early 2000's revealed that atomic and H$_2$ jets with projected extensions on the plane of the sky larger than one parsec (so-called {\it parsec-scale jets}) are a common phenomenon
\citep[e.g.][]{1997AJ....114..280E, 1997AJ....114.2708R, 1999MNRAS.310..331M, 2000A&A...354..236E, 2000A&A...355..639S, 2004A&A...415..189M, 2004A&A...420..975M}.
More recent cloud-wide surveys continue to find new giant jets, 
indicating that 
young stars of all masses can power flows that interact with their surroundings 
at parsec-scale distances \citep[e.g.][]{2008MNRAS.387..954D, 
2009A&A...496..153D, 2012AJ....144..143B, 2012MNRAS.425.1380I}.
In many cases these wide-field observations reveal that jets 
originally thought to extend less than about 0.5 pc, in reality extend 2 to 3 pc (or even more) on the sky.

The fact that parsec-scale protostellar jets 
are a common phenomenon should not have come as a surprise since, 
assuming (constant) jet velocities of 100 to 300~km~s$^{-1}$ and timescales of at least $2 \times 10^5$~yr 
(the approximate lifetime of the Class I stage), their expected size would be (at least) 20 to 60 pc. Even when deceleration of the ejecta is considered, they are expected to reach sizes of a few pc at an age of $\sim 10^4 - 10^5$ yr \citep{2000A&A...354..667C, 2004ApJ...608..831G}. Hence, parsec-scale protostellar flows should be a common phenomenon (if not the norm). 

In many cases, however, these giant jets 
have been hard to detect in the optical and NIR as very wide-field images are needed to cover their entire extent. Moreover, jet time variability leads to gas accumulation in knots (working surfaces) with low density between them. Thus giant HH flows do not show a continuous bright emission, (unlike, for example, microjets or HH jets within a few 0.1 pc of their powering source which exhibit closely spaced bright knots arising from shorter modes of variability; see Sect.~\ref{sec:precess}). 
Instead, giant jets appear as a sparse chain of diffuse and fragmented HH or H$_2$ knots separated by distances of 0.1 pc to 2 pc. Without proper motion studies, it is sometimes hard to distinguish between knots from different jets and to properly identify their source.

Millimeter CO observations have shown that giant jets can entrain the ambient molecular gas and produce large ($> 1$ pc), massive (a few solar masses or more) bipolar shells of swept-up gas at medium velocity ($\simeq$ 10 km/s), often referred to as ``(giant) molecular outflows"
\citep[e.g.][]{1997ApJ...491..653T, 2001ApJ...554..132A, 2002ApJ...575..911A, 2006ApJ...649..280S}.
These observations show that even when they are too narrow to fully disperse the dense envelope around their source, jets can impact the density and kinematic distribution of their (less dense) parent clump and cloud, out to distances greater than a parsec away from the source. 

\begin{figure}[t]
 \epsscale{0.95}
 \plotone{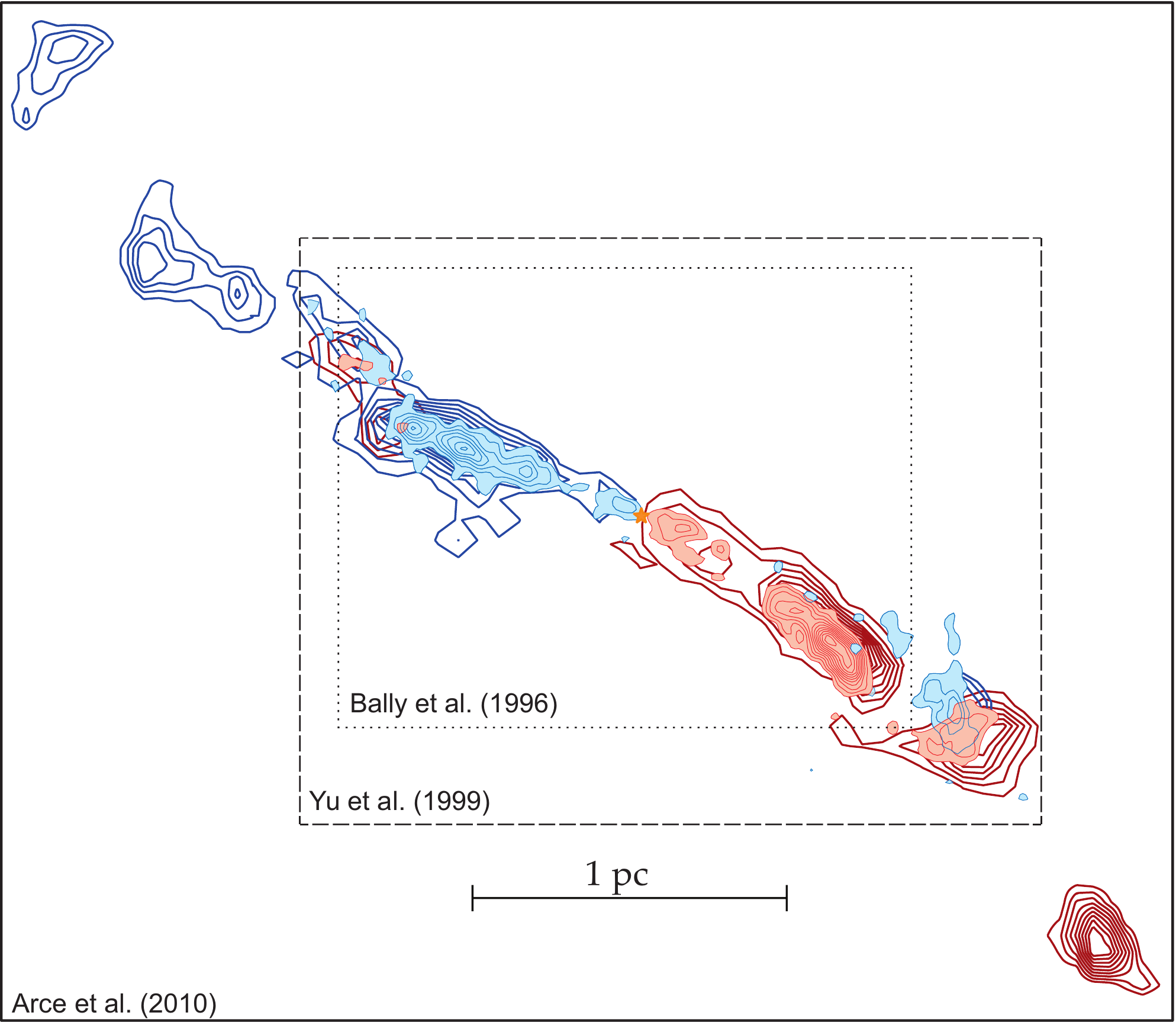}
  \caption{\small Our changing view of the size of the giant molecular outflow from the Class I source B5-IRS1 (orange star symbol at center). The dotted square shows the original extent from \citet{1996ApJ...473..921B}. The dashed square shows the region mapped by \citet{Yu1999}. The dark blue/red contours show the map from the cloud-scale CO outflow survey of \citet{2010ApJ...715.1170A}.
 }  
 \label{arce}
\end{figure}

Recent cloud-wide CO maps, like optical and IR surveys, have  shown that molecular outflows can be much larger than previously thought (see Fig. \ref{arce}), and have helped increase the number of known giant flows 
\citep{2007ApJ...660..418S, 2010ApJ...715.1170A, 2012MNRAS.425.2641N}.
For example, in an unbiased search using a cloud-wide CO map of Taurus,  \citet{2012MNRAS.425.2641N} found that 40\% of the twenty detected outflows have sizes larger than one parsec. Given the difficulty in detecting the entirety of giant outflows (see above), it would not be surprising if most outflows from late Class 0 sources (and older) have scales of a parsec or more.

We note that many giant HH jets extend beyond the confines of their parent molecular cloud. Thus an observed molecular outflow only traces the swept-up gas lying within the molecular cloud: see for example, HH 111 \citep{1996ApJ...460L..57C,2007ApJ...658..498L},  HH 300 \citep{2001ApJ...554..132A}, HH 46/47 \citep{2009A&A...501..633V, 2013ApJ...774...39A}. This implies that giant HH jets most likely drive {\em atomic hydrogen outflows} in the intercloud medium as well, and may also be a source of turbulence in the low density (atomic) ISM. Future galactic HI interferometric surveys should help assess the impact of giant outflows on the atomic medium. 

\subsection{Fundamental Issues In Jet/Outflow Feedback on Clouds}
Outflow feedback touches on two critical issues facing modern theories of star formation; the relative inefficiency of star formation, and the origin of turbulence in clouds \citep{2007ARA&A..45..565M, 2004ARA&A..42..211E}. The first issue relates to the fact that observations find surprisingly low values of the star formation efficiency ($SFE$) in clouds, with typical values ranging from 0.01 to 0.1.  Theoretical accounts for the low values of $SFE$ rely on some form of support such as supersonic turbulence within the cloud to keep it from collapsing. But while turbulence can provide an isotropic pressure support, both hydrodynamic and MHD turbulence decay quickly \citep{1999ApJ...524..169M, 1998ApJ...508L..99S}.  Thus turbulent motions must be continually driven, either internally via gravitational contraction 
and stellar feedback, or externally via turbulence in the general ISM if clouds are long-lived. How this ``driving" takes place and (self-)regulates the SFE is the second critical issue.

Thus a fundamental question facing studies of both turbulence and star formation efficiency is the role of stellar feedback, via both radiation and outflows.  The various feedback mechanisms in star formation are reviewed in the chapter by Krumholtz et al. and in \citet{2011IAUS..270..275V}. Here we focus on {\em outflow-driven} feedback, and the circumstances under which it could {\it substantially} change conditions in a star-forming cloud.  With respect to turbulence, 
the question becomes: (a) do protostellar outflows inject enough momentum to counteract turbulence decay in clouds; (b) can outflows couple to cloud gas on the correct scales to drive turbulent (rather than organized) motions.

Answering question (a) requires that a steady state can be established between the dissipation rate of the turbulent momentum in a cloud ($dP_{\rm turb}/dt$) and the momentum injection rate by protostellar outflows, ($dP_{\rm out}/dt$).  When such a steady state is achieved the cloud is close to virial equilibrium  (Fig. \ref{Hartigan}).  Note that the dissipation rate of turbulent momentum can be written as a function of the cloud mass $M_{\rm cl}$, its turbulent velocity dispersion $V_{\rm vir}$ and the dissipation time $t_{\rm diss}$ as \citep{2011ApJ...740...36N}
\begin{equation}
\frac{dP_{\rm turb}}{dt} = \alpha \frac{M_{\rm cl} V_{\rm vir}}{t_{\rm diss}}
\end{equation}
where $\alpha$ is a factor close to unity. The outflow momentum injection rate can be written as 
\begin{equation}
\frac{dP_{\rm out}}{dt} = \epsilon_{\rm SFR} \times f_{\rm w} V_{\rm w} \ ,
\label{eq:outflow}
\end{equation}
where  $\epsilon_{\rm SFR}$ is the star formation rate in solar masses per year, $f_{\rm w}$ is 
fraction of stellar mass injected as wind and $V_{\rm w}$ is the wind velocity. 
As we will see, both observations and simulations suggest that on {\em cluster scales} a balance between these terms can be achieved.  Thus the implication is that outflow momentum deposition is sufficient to 
lead to the observed values of $\epsilon_{\rm SFR}$. 

It is worth noting that essential elements of the problem can be captured via dimensional analysis \citep{2007ApJ...659.1394M}.  By considering a cloud of mean density $\rho_0$, with outflows occurring at a rate per volume $\mathcal{S}$ and with momentum $\mathcal{I}$ one can define characteristic outflow scales of mass, length, and time:
\begin{equation}
\mathcal{M}=\frac{\rho_0^{4/7}\mathcal{I}^{3/7}}{\mathcal{S}^{3/7}},
\mathcal{L}=\frac{\mathcal{I}^{1/7}}{\rho_0^{1/7}\mathcal{S}^{1/7}},
\mathcal{T}=\frac{\rho_0^{3/7}}{\mathcal{I}^{3/7}\mathcal{S}^{4/7}}
\end{equation}
Combining these gives other characteristic quantities.  Of particular interest is the characteristic
velocity:
\begin{equation}
\mathcal{V}=\frac{\mathcal{L}}{\mathcal{T}}=\frac{\mathcal{I}^{4/7}\mathcal{S}^{3/7}}{\rho_0^{4/7}}=\frac{\mathcal{I}}{\rho_0
\mathcal{L}^3}
\end{equation}
Assuming typical values for $\rho_0$, $\mathcal{I}$, and $\mathcal{S}$ in cluster environments yields a supersonic characteristic mach number of $M =\mathcal{V}/c > 1$.   This suggests that outflows contain enough momentum to drive supersonic turbulence. Note however that these relations are for spherical outflows and an open question relates to how more narrow bipolar outflows will couple to the cluster/cloud gas. Jet wandering may play a key role here. One must also be careful when measuring the typical momentum in outflows $\mathcal{I}$ to account for non-emitting molecular gas (due to dissociation) and to very low velocity gas representing decelerating fossil cavities just before they are subsummed by background turbulence. We will discuss this issue in Section 4.4

The question of outflow feedback altering the star forming properties of a cluster/cloud is a more complex issue as there are a number of ways to characterize the problem.  Outflows can directly alter $SFE$ by providing turbulent support against gravity as discussed above, or they may help unbind gas from either individual cores or the cluster environment as a whole. In addition, outflows could change the global properties of star formation by starving still-forming higher mass stars of their reservoirs of gas and therefore shifting the mean stellar mass of a cluster to lower values.

Note that we have been careful to distinguish between feedback on clump/cluster scales and that on scales of the larger parent clouds. Outflows likely represent the lowest rung of a ``feedback-ladder"; a sequence of ever-more-powerful momentum and energy injection mechanisms that operate as more massive stars form.  {\em Thus collimated outflow are likely to be most effective in driving feedback on cluster rather than full GMC scales}. We note however that magnetic fields may provide more effective coupling between outflows and larger cloud scales \citep{2005MNRAS.359..164D}.


\subsection{Observations of jet-cloud interactions: Momentum Budget and Turbulence Driving Scale}

\begin{figure}[t]
 \epsscale{1.0}
 \plotone{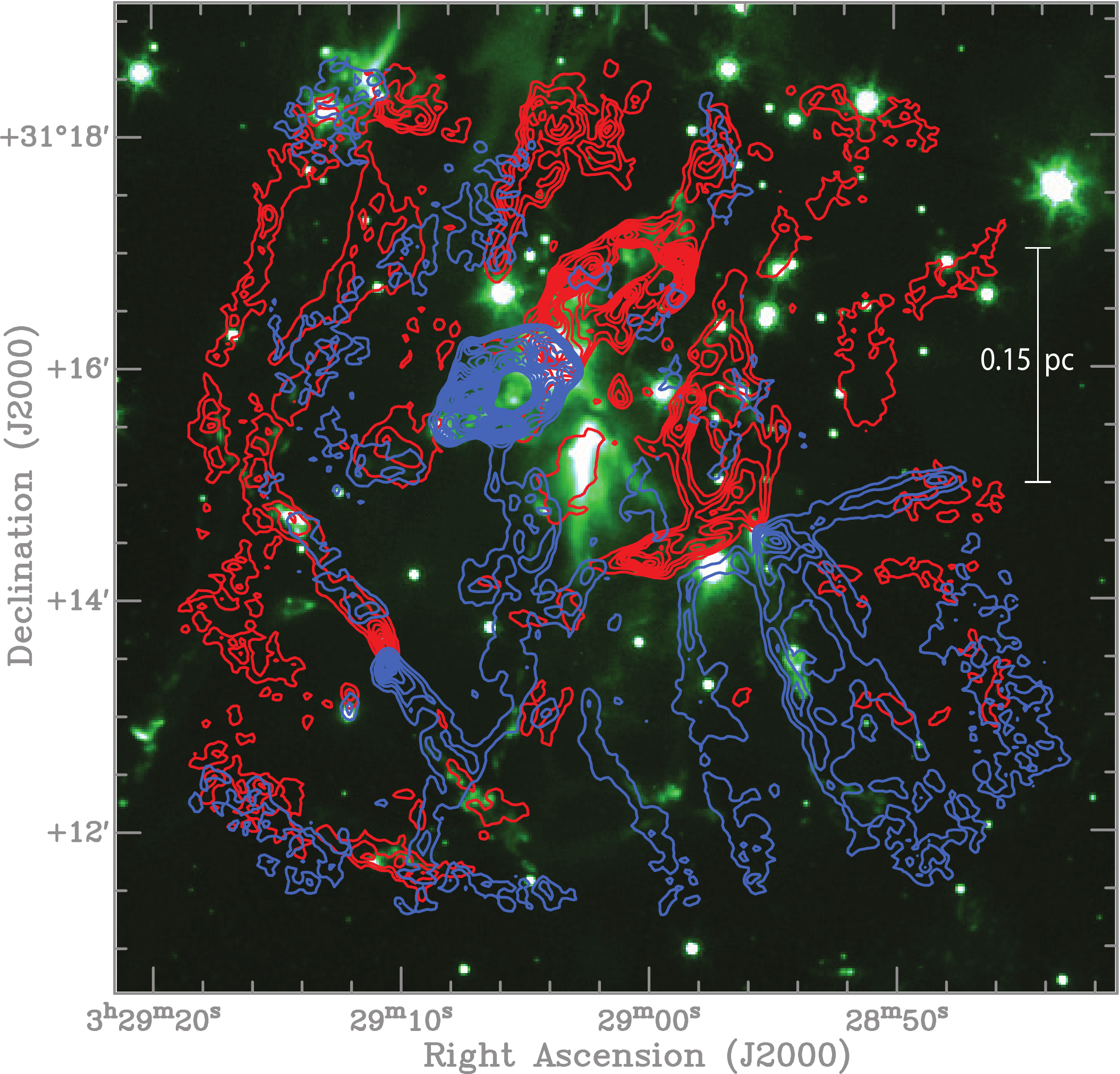}
 \caption{\small Map of molecular outflows in the central region of the protostellar cluster NGC 1333, overlaid on a map of the IRAC 4.5 $\mu$m emission.  Blue and red contours show the integrated intensity of the CO(1-0) blueshifted and redshifted outflow emission, from the CARMA interferometric observations by Plunkett et al. (2013)} 
\label{Eisoeffel}
\end{figure}

Numerous attempts have been made to investigate observationally how much feedback protostellar outflows are providing to their surrounding cloud (Fig \ref{Eisoeffel}). We note that in the line of virial analyses, many such studies have focussed on comparing the kinetic {\em energy} in outflows to that in cloud turbulence or gravitational binding; but since energy is not conserved (because of strong radiative losses) as the ouflow sweeps up mass and eventually slows down to merge with the background cloud, it is very important to put more emphasis on the measurement of the outflow {\it momentum}, a conserved quantity, to come to meaningful conclusions. We will still quote the energy budgets here for completeness, but will focus on the relevant momentum estimates to reach our conclusion. 

\citet{2010MNRAS.409.1412G} present various CO line maps of the
Serpens molecular cloud obtained in the course of the JCMT Gould Belt Legacy
Survey. Because of the complexity of spatially overlapping outflows in this
crowded star forming region the analysis of outflow properties is based on the
blue/red-shift deviation of the $^{12}$CO velocity with respect to C$^{18}$O. 
The latter is optically thin across the cloud, does not trace outflows, and thus defines
a kind of local rest velocity. After correction for a random inclination
distribution of the flows, this study finds the total outflow energy to be
approximately 70\% of the total turbulent energy of the region. 
Similar conclusions have come from studies of other regions such as 
$\rho$ Ophiuchi \citep{2011ApJ...726...46N}, Serpens South \citep{2011ApJ...737...56N},  
L1641-N \citep{2012ApJ...746...25N} and NGC2264C \citep{2009A&A...499..175M}.

\citet{2010ApJ...715.1170A, 2011ApJ...742..105A} analyzed the COMPLETE CO
datasets of the entire Perseus star forming complex to find new outflows, and wind-driven shells around
more evolved Class II stars resulting from the interaction of wide-angle winds with the
cloud material. This study more than doubled the amount of outflowing mass,
momentum, and kinetic energy of protostellar  outflows in Perseus. They calculate that the total
outflow kinetic energy in the various star forming regions within the Perseus
cloud complex (e.g. B1, B5, IC348, L1448, NGC1333) amounts to about 14-80 \% of the local total turbulent energy, and to 4 to 40 \% of the total
gravitational binding energy in these regions.
In the same regions, the total outflow {\em momentum} is typically 10\% of the cloud turbulent momentum (up to 35\% in B5). If one takes into account that these outflows most likely have ages of only 0.2 Myrs
or less, while molecular clouds have lifetimes of about 3 -- 6 Myrs
(e.g. \citet{2009ApJS..181..321E}) it becomes clear that a few generations 
of outflows will suffice to provide a very significant source of momentum input to each cloud. With the flow timescales and turbulence dissipation timescales estimated by \citet{2010ApJ...715.1170A}, the mean rate of momentum injection by outflows is only a factor 2.5 less than the rate of turbulent momentum dissipation. We note that a comparable fraction of outflow momentum
could be hidden in the form of atomic gas, swept-up and dissociated in shocks faster than 25 km/s \citep{2007A&A...471..873D}. 
Taken together, these observations indicate that directed momentum injection by outflows could significantly contribute to sustaining observed levels of turbulence. Similar detailed studies of other regions are certainly necessary in the future to quantify the impact of jets and outflows on their surrounding clouds. 

This raises the question of whether outflows are also able to ultimately disrupt their cloud by dispersing and unbinding cloud material. \citet{2010ApJ...715.1170A} find that outflows in Perseus currently carry momentum enough to accelerate only 4 to 23 per cent of the mass of their respective clouds to the local escape velocity. But multiple generations of outflows will again increase their impact. Hence it is clear that outflows are within range of unbinding some fraction of their parent clusters. A plausible scenario proposed by \citet{2010ApJ...715.1170A} would be that outflows help disperse a fraction of their surrounding gas and other mechanisms, such as dispersion by stellar winds and erosion by radiation, help dissipate the rest of the gas that does not end up forming stars. 

Observational studies have also been used to estimate the role played by outflows in the injection of cloud turbulence.  These efforts consist in attempts to constrain the scale at which turbulence is driven into molecular clouds. 
\citet{2009A&A...504..883B} and \citet{2009ApJ...707L.153P} conducted such studies on the NGC 1333 star forming region using principal component
analysis (PCA) in the former case and velocity component analysis (VCS) in the latter. In both cases, analysis of CO line
maps from the COMPLETE survey \citep{2006AJ....131.2921R} 
were compared with a corresponding analysis on synthesized maps from numerical
simulations of clouds with turbulence driven at various scales (in Fourier space). Both find that the observations are only consistent with simulated turbulence driven at large scales on the size of the entire NGC 1333 region. Thus they come to the conclusion that turbulence should mostly be driven externally, and that outflows --- as small-scale driving sources within the molecular cloud --- should not play a major role. 

This appears to contradict the above observations indicating that outflows are a major source of momentum input, at the cluster levels at least. However, \citet{2010ApJ...715.1170A} comment
that simulations of turbulence in Fourier space, with a necessarily limited
range of wave numbers, may cause a difference between flows as they appear in these simulations and turbulence in nature. Simulations of outflow-driven feedback do lend support to this interpretation \citep{2010ApJ...722..145C}, and we come back to this crucial issue in Section 4.6.

\subsection{Physical Processes In Jet-Cloud Feedback}

Propagating jets can, in principle, entrain environmental material through two (not totally unrelated) processes. First there is {\it Prompt Entrainment} which is the incorporation of material in shocks such as the working surface at the leading head of a jet, or internal working surfaces produced by a time-dependent ejection \citep[see][]{1993ApJ...414..230M, 1993A&A...278..267R}. A second mechanism is {\it Side Entrainment} which is the incorporation of material through a turbulent mixing layer at the outer edge of the jet beam \citep{1993A&A...276..539R}. While both processes are likely to shape the interaction of jets with their environments, prompt entrainment is likely to be more important for feedback on cluster and cloud scales since it will often be the fossil swept-up shells (bounded by shocks) which couple outflow momenta to the cloud.

The jet-to-cloud momentum transfer efficiency varies inversely with the jet-to-cloud density ratio \citep{1993ApJ...414..230M}. An overdense jet will ``punch" fast through the cloud without depositing much momentum into the swept-up shell. The efficiency will increase  when a jet impacts a {\em denser} region of the molecular cloud (e.g., a molecular cloud core). In such an interaction, the jet will initially be deflected along the surface of the dense core, but at later times the jet will slowly burrow a hole into the core \citep{2002eaa..bookE1910R}. During this burrowing process, most of the momentum of the jet is transferred to the cloud core material. 

\begin{figure}[t]
 \epsscale{0.85}
 \plotone{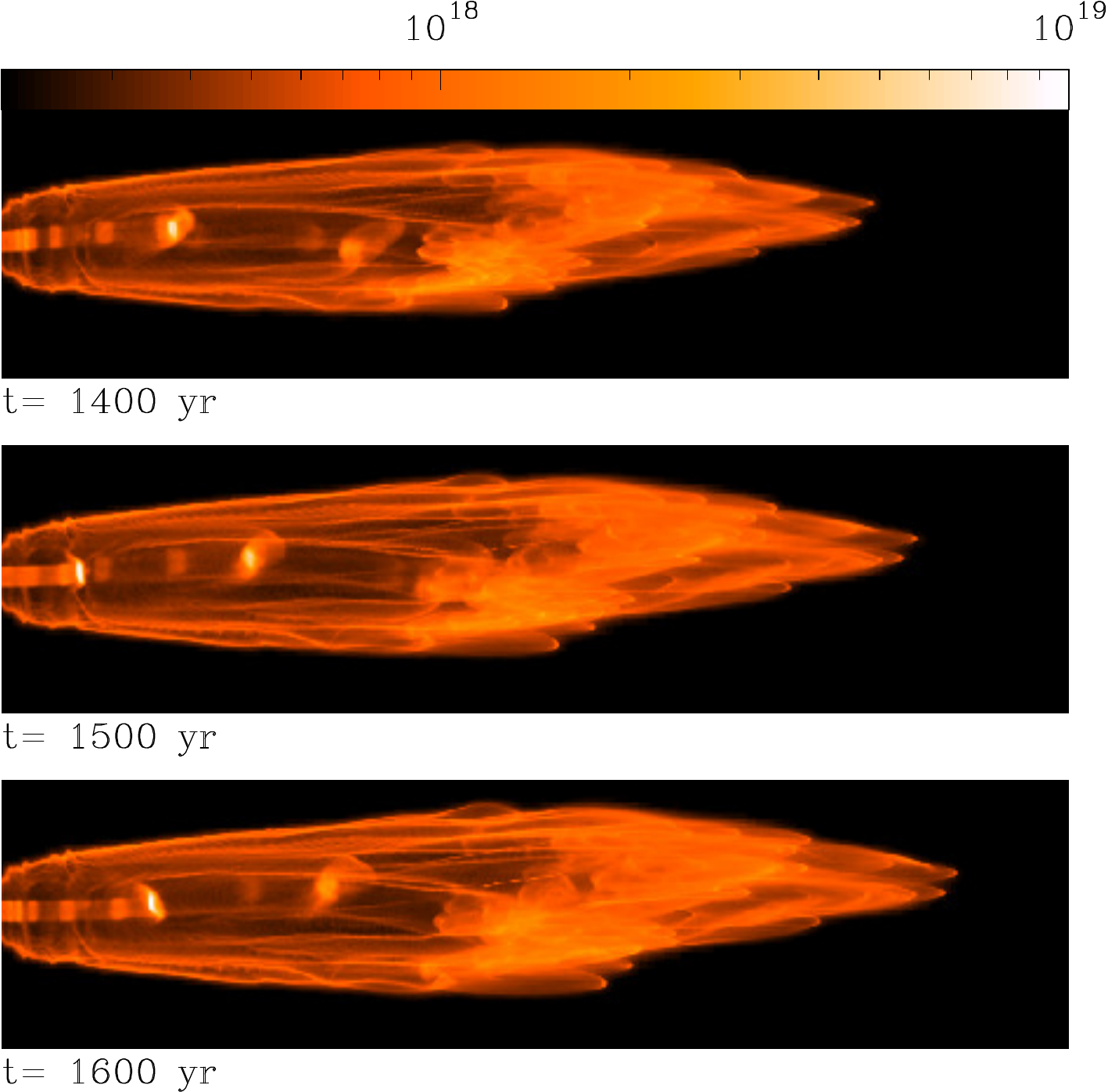}
 \caption{\small Column density time-sequence computed from a model of a jet with
a variable ejection velocity and a precession of the outflow axis.
The initial jet radius is resolved with 10 grid points
at the highest resolution of a 5 level adaptive grid.}  
\label{raga}
\end{figure}

Efficient momentum deposition also occurs if the jet ejection direction is time-dependent (due to precession of the jet axis or orbital motion of the source, see Section 3 and \citep{2009ApJ...707L...6R}).  This will be particularly true when a variable jet direction is combined with a variable ejection velocity modulus (Fig \ref{raga}). These effects break the jet into a series of ``bullets'' travelling in different directions \citep{1993MNRAS.264..758R}. Alternatively, the ejection itself might be in the form of discrete ``plasmoids'' ejected along different paths \citep{2008ApJ...672..996Y}. These bullets differ from the leading head of a well aligned jet in that they are not resupplied by material ejected at later
times. Therefore, they slow down as they move through the molecular cloud due to ram pressure braking as seen in the giant HH34 jet complex \citep{2000A&A...354..667C, 2002ApJ...580..950M}. Most of the jet momentum could then be deposited within the molecular cloud, instead of escaping into the atomic ISM. 

When such ``bullets'' eventually become subalfv\'enic and/or subsonic (depending on the cloud magnetization) their momentum can be efficiently converted into MHD wave-like motions of the molecular cloud \citep{2005MNRAS.359..164D}. This is true for any form of decaying jet flow (mass-loss decreasing over time).  Thus
"fossil'' swept-up shells that expand and slow down after the brief Class 0 phase should be the main agents coupling outflow momentum to cloud turbulence. Numerous such ``fossil cavities'' have been discovered in regions such as NGC 1333 \citep{2005ApJ...632..941Q}, and across the Perseus cloud \citep{2010ApJ...715.1170A,2011IAUS..270..287A}. Observationally derived scaling properties for momentum injection in such flows \citep{2005ApJ...632..941Q} have been recovered in simulations of shells driven by ``decaying'' outflows \citep{2009ApJ...692..816C}. 

The direct turbulent driving and/or coupling of {\it individual} outflows to the cloud has been investigated numerically by a number of authors. While \citet{2007ApJ...668.1028B} showed that a single active outflow (ie a Class 0 source) in a quiescent medium would not drive turbulent motions, \citet{2009ApJ...692..816C} demonstrated that a fossil outflow in an already turbulent cloud will fragment, and re-energize those turbulent motions. This speaks to the complex issue of ``detrainment'' (i.e., the eventual merging of material from the outflow into the surrounding environment).  While it is clear that jets can re-energize turbulence, the end states of detrainment remains an important issue needing resolution in order to calculate the full ``feedback'' of the momentum provided by jets into the turbulent motions of the placental molecular cloud. 

Most importantly, large-scale simulations (discussed in the next section) show that interactions (collisions) between {\em multiple} outflows on scale $\mathcal{L}$ (from equ. 3)
may be the principle mechanism for converting directed outflow momentum into random turbulent motions. Conversely, the role of cloud turbulence in altering outflow properties was explored in \citet{2011ApJ...743...91O}.  In that study, turbulent motions associated with collapse produced asymmetries between the red and blue swept-up outflow lobes.  This study also showed that some caution must be used in converting observations of outflows into measurements of injected momentum, as low-velocity outflow material can be misidentified as belonging to turbulent cores. 


\subsection{Large-scale simulations of outflow feedback}

\begin{figure}[t]
 \epsscale{.85}
 \plotone{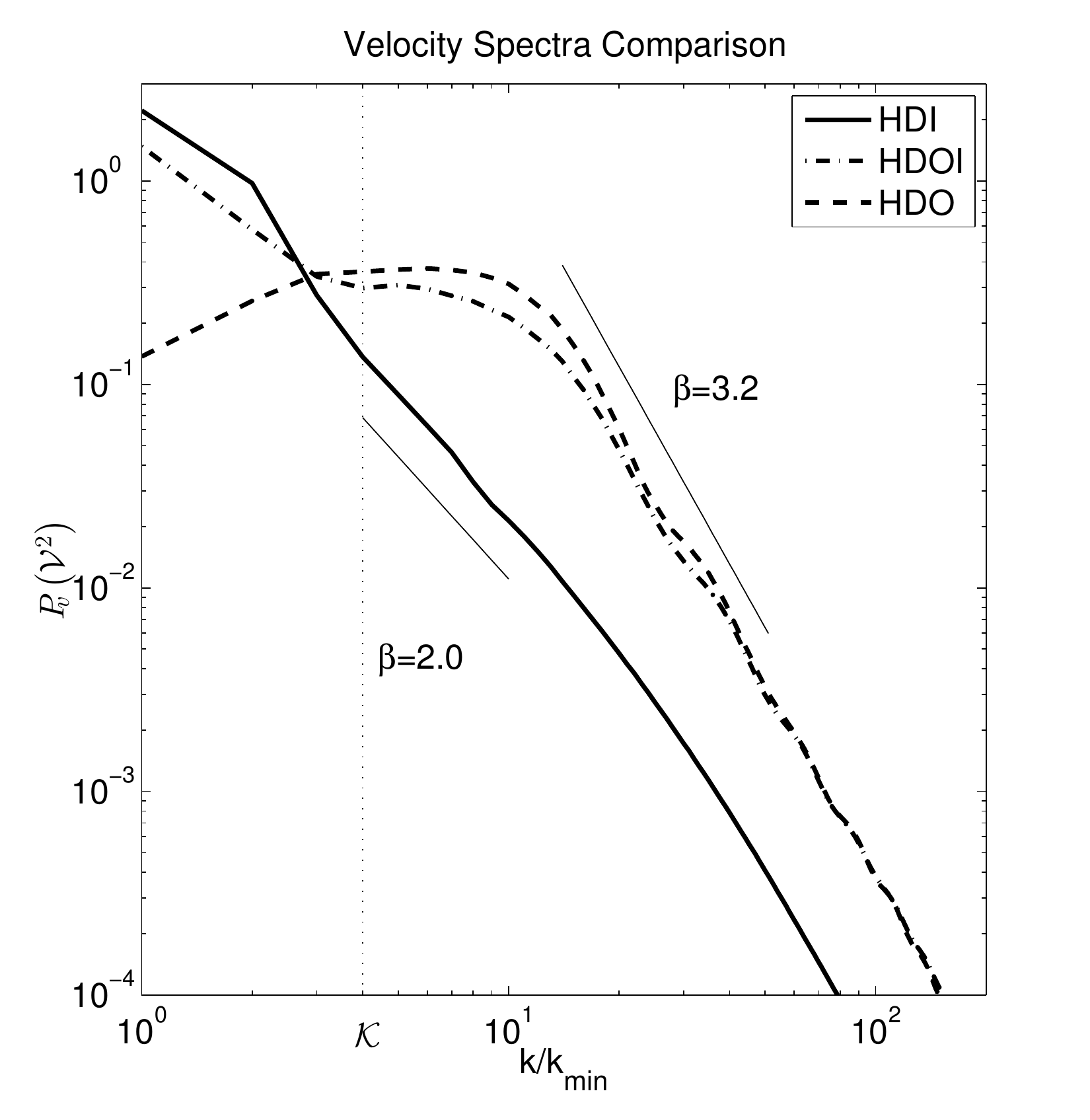}
 \caption{\small Velocity power spectrum for runs with pure fourier driving (HDI, solid line), fourier+outflow driving (HDOI, dash-dotted line), and pure outflow driving (HDO, dashed line) from simulations by \citet{2010ApJ...722..145C}.  The vertical dashed line corresponds to the outflow interaction wave number 
 $\mathcal{K} = 1/\mathcal{L}$. Dotting lines show $k^{-\beta}$ for $\beta = 2$ and $\beta = 3.2$.} 
 \label{frank}
\end{figure}

Analytic models such as those of \citet{2000ApJ...545..364M} and \citet{2007ApJ...659.1394M} have
articulated basic features of outflow driven feedback such as the scaling laws discussed in 
section 4.3.  The inherently three-dimensional and time-dependent nature of outflow feedback, 
however, requires study through detailed numerical simulations 
\citep[see also the review by][]{2011IAUS..270..275V}.  

The ability of multiple outflows to generate turbulent support within a self-gravitating clump was first addressed in the simulations of \citet{2006ApJ...640L.187L} and \citet{2007ApJ...662..395N}, which started with a centrally condensed turbulent clump containing many Jean's masses. The initial turbulence generated overdense regions which quickly became Jeans unstable, initiating local regions of gravitational collapse. Once a density threshold was crossed, these collapsing regions were identified as protostars. Mass and momentum was then driven back into the grid in the form of outflows.  

The most important conclusion of these studies was that once star formation and its outflows commenced, the young cluster achieved a dynamic equilibrium between momentum input and turbulent dissipation.  It is noteworthy that these studies found bipolar outflows more effective than spherical winds for turbulent support, as the former could propagate across longer distances. Star formation efficiencies of just a few percent were achieved in simulations with outflow feedback.

Another key point to emerge from simulations of outflow-feedback is the nature of the turbulence it produces.  In \citet{2007ApJ...662..395N} a break in the velocity power spectra $E(k)$ was identified, below which (i.e. longer scale-lengths) the spectrum flattened.  This issue was addressed again in \citet{2009ApJ...695.1376C} who ran simulations of the interaction of randomly oriented interacting bipolar outflows. In this work, the outflow momentum injection rate was made time-dependent to explore the role of fossil shells in coupling to the cloud turbulence.  \citet{2009AAS...21442506C} also found a well defined ``knee'' in the spectrum at $\mathcal{K} \propto 1/\mathcal{L}$, the interaction scale defined via dimensional analysis (see eq. 3). Thus, the collision of fossil outflows cavities and the subsequent randomization of directed momenta was responsible for generating the observed turbulence.   

\citet{2009AAS...21442506C} also found that outflow-driven turbulence produced a power spectrum that steepened above the knee as $E(k) \propto k^{-3}$ (see Fig. \ref{frank}). In contrast, standard turbulence simulations using forcing in Fourier space typically find ``Burger's'' values of $k^{-2}$. The steeper slope was caused by outflow shells sweeping up eddies with wavenumbers higher than $\mathcal{K}$.  The presence of both a knee and a steeper slope in the spectrum of outflow-driven turbulence offers the possibility for observation of these, and perhaps other, signatures of outflow feedback. Note that changes to the turbulent spectra via outflows remained even in the presence of driving at scales larger than 1/$\mathcal{K}$ \citep{2010ApJ...722..145C}.  Modifications of density probability distribution functions (PDFs) of the ambient medium via outflow-driven turbulence were also reported in \citet{2013MNRAS.432L..80M}. 

PCA methods applied to datacubes from simulations of outflow-driven turbulence demonstrate that the discreet, small scale sources can artificially appear overwhelmed by larger scale flows, even if those flows have far less power \citet{2010ApJ...722..145C}.  From these results it is likely that the issue of observational determination of the correct driving scale(s) of turbulence remains an open question.  Note that the issue is not just the largest scales at which driving occurs, but which process dominates on the scales where star formation occurs.  Thus even if turbulence cascades down from GMC scales, outflow feedback on cluster scales may still be important in determining local star formation efficiencies and related properties.

Because magnetic fields are closely tied to the origin of protostellar outflows, exploring the combined role of magnetic field and outflows feedback has been an important issue.  Using AMR methods, \citet{2010ApJ...709...27W} began with a turbulent, moderately condensed clump of $\sim 1600 M_\circ$ and found that in the absence of regulation by magnetic fields and outflow feedback, massive stars would readily form within a cluster of hundreds of lower mass stars.  These simulations showed that the massive stars were fed by material infalling from large scales (i.e. clump-fed rather than core-fed accretion).  The importance of large scale accretion modes made high mass star formation particularly susceptible to disruption by outflows.  Once mass loss was initiated by lower mass stars, their outflows eroded the dense filaments feeding massive star formation. In addition, at later times the induced turbulent motions of interacting outflows slowed down the global collapse modes that had continued to fuel the young massive stars. Thus \citet{2010ApJ...709...27W} found global accretion rates were reduced, leading to fewer high-mass stars by the simulations end. \citet{2010ApJ...709...27W} and \citet{2011ApJ...740...36N} also found strong links between outflow feedback and magnetic fields.  Even an initially weak field could retard star formation as the field was amplified to equipartition strength by the outflow-driven turbulence, with the "turbulent" field component dominating the uniform one

Using AMR methods, \citet{2012ApJ...747...22H} studied low-mass star formation in the presence of outflow {\it and} radiation feedback.  These simulations found that outflows reduce protostellar masses and accretion rates by a factor of three each. In this way, outflows also led to a reduction in protostellar luminosities by an order of magnitude. This reduced the radiation feedback, and enhanced fragmentation.  In contrast with previous results, \citet{2012ApJ...747...22H} found that the outflows did not change the global dynamics of the cloud because they were narrow and did not couple well to the dense gas. \citep{2012ApJ...754...71K} studied the role of outflow (and radiation) feedback in high-mass star forming regions. Their results also indicated a smaller impact from outflows.  Note that both these these simulations did not include magnetic fields.

Finally we note that almost all simulations of outflow-feedback rely on common parametrizations of the individual outflows.  In particular the total outflow momentum is expressed as $P_o = f_w V_w M_*$ making the combination $f_w V_w$ the outflow momentum per unit stellar mass.  In their analytic description of outflow feedback and star formation, \citet{2000ApJ...545..364M} assumed a value of $f_w V_w = 40$ km/s.  A review of the literature yields an observational range for this parameter of $10 < f_w V_w < 25$ km/s \citep{1997A&A...323..549H, 1996A&A...311..858B, 2000prpl.conf..867R, 2013ApJ...774...22P}.  Assuming $V_w = 100$ km/s yields $0.01 < f_w < 0.25$.  The presence of a wide-angle wind relative to a collimated flow component is another key parameter used in simulations, sometimes expressed as the ratio of momentum in a fully collimated component to a spherical one, $\epsilon = P_c/P_s$. To the extent that observations provide a guide for this parameter, it would appear that $\epsilon \gg 1$ is favored, since wide-angle winds do not appear to carry as much momentum as the jets (at least in the class 0 phase, where most of the momentum is injected).  Finally we note that care should be taken in how "outflow" momentum is added to the grid in feedback simulations.  Given that the large speeds associated with the winds can slow down simulations (via CFL conditions) momentum is sometimes added via lower-speed higher-mass flows, or given directly to ambient material in the vicinity of the source.  Further work should be done to test the effect these assumptions have on turbulence injection and outflow feedback on star formation. 

\section{\textbf{Explosive Outflows}}

Finally we note another class of mass-loss may play an important role in delivering momentum back into the parent cloud, i.e. explosive though non-terminal (ie. non-supernova) outflows. An archetype of this phenomena is the BN/KL region in Orion which produced a powerful ($\sim 10^{47}$ to $^{48}$ erg) wide-angle explosion  approximately 500-1000 yrs ago \citep{1993PASAu..10..298A, 2002AJ....124..445D, 2011ApJ...727..113B, 2011ApJ...739L..13G}. The origin of the outflow appears to lay in a non-hierarchical multiple star system that experienced a dynamical interaction leading to the ejection of 2 members and the formation of a tight binary or possibly a merger \citep{2005ApJ...627L..65R, 2005AJ....129.2281B, 2005ApJ...635.1166G, 2008ApJ...685..333G, 2009ApJ...704L..45Z, 2011IAUS..270..247B}. Proposed scenarios for powering the outflow involve the release of energy from envelope orbital motions, gravitational binding of the tight pair, or magnetic shear. The rapid release of energy leads to the fastest ejecta emerging from deep within the gravitational potential of the decaying cluster. Rayleigh-Taylor instabilities are then triggered as this material plows through a slower-moving, previously ejected envelope.  The fragmented ejecta which are created will be effective at driving turbulence in their surroundings, like the ''bullets" in precessing jets.     

If such a mechanism operates in other massive star forming regions, it may be an important source of outflow feedback. The {\em Spitzer} Space Telescope detected at 4.5 $\mu$m a wide-angle outflow similar to BN/KL in the $10^6 L_\odot$ hot core G34.25+0.16 (located at 5kpc in the inner Galaxy,\citep{2008AJ....136.2391C}). Source G in W49, the most luminous water maser outflow in the Milky Way, may be yet another example \citep{2009MNRAS.399..952S}. Finally \citet{2008ApJ...680..483S} found evidence for interstellar bullets having a similar structure to the BN/KL ''fingers'' in the outflow from the massive young protostar  IRAS 05506+2414.   

\section{Conclusions and Future Directions}
In this chapter we have attempted to demonstrate that protostellar jets and outflows are not only visually beautiful and important on their own as examples of astrophysical magneto-fluid dynamical processes, but they are also an essential player in the assembly of stars across a remarkable range of size-scales. We find that issues of feedback from jets/outflows back to the star formation process is apparent on three scales: those associated with planet formation, those associated with the natal core, and those associated with clustered star formation.

On scales associated with planet-forming disks, jets can impact planet assembly through disk irradiation/shielding and MHD effects associated with outflow launching.  These processes can alter disk properties in those regions where planets will be forming. Future work should focus on articulating the feedback between jet driving and the disk mechanisms associated with creating planets. 

On the scales of the natal cores, jets and outflows seem capable of explaining the low observed 30\% core-to-star efficiency, through a combination of powerful MHD ejection during the earliest collapse phase, and envelope clearing by wide-angle winds during the later phases.

On scales associated with clusters (or perhaps clouds) multiple jets/outflows can drive turbulence, alter star formation efficiencies and affect the mean stellar mass. It is also possible that outflows may help unbind cluster gas.  Future work should focus on providing simulations with the best parametrizations of jet properties (such as momentum injection history and distributions), characterization of physical mechanisms for feedback, and exploration of feedback observational signatures. The full range of scales over which both collimated and uncollimated outflows from young stars impact the star-formation process must also be articulated.  Observational signatures of such feedback must also explored.

With regard to the physics of jets themselves, new results make it clear that they are collimated magnetically on inner disk scales, and include multiple thermal and chemical components surrounded by a {\em slower} wide-angle wind. The emerging picture is that of a powerful MHD disk wind, collimating the inner stellar wind and magnetospheric ejections responsible for braking down the star. Detailed analysis and modeling is needed to confirm this picture and articulate the properties of these different components as a function of source age.

We have also shown how High Energy Density Laboratory Astrophysics (HEDLA) experiments have already contributed new and fundemnetal insights into the hydrodynamic and MHD evolution of jets.  We expect HEDLA studies to grow beyond jet research in the future as they hold the promise of touching on many issues relevant to star and planet formation (i.e. cometary globules and hot Jupiters).

Observationally we expect new platforms to hold great promise for jet and outflow studies.  In particular ALMA and NIR IFUs should prove crucial to resolving jet rotation profiles, shocks and chemical stratification in statistically relevant jet samples, and to better understand their interaction with the surrounding envelope. Such data will provide definitive tests of disk wind models.  NIR interferometry of CTTS (eg. with GRAVITY on VLTI) promises to be a powerful test of atomic jet models. Synchrotron studies with eVLA, LOFAR should allow jet magnetic fields to finally come into view.  Finally, long baseline monitoring of the short quasi-periodic knot modulation in jets ($\sim 3-15$ yrs) should allow to clarify the origin of these features and their link with stellar and disk physics (magnetic cycles, accretion outbursts) and source binarity.

\textbf{Acknowledgments} Frank acknowledges support from the NSF, DOE and NASA. Ray acknowledges support from Science Foundation Ireland under grant 11/RFP/AST3331. Cabrit acknowledges support from CNRS and CNES under the PCMI program. Arce acknowledges support from his NSF CAREER award AST-0845619.

\bigskip
\parskip=0pt

\bibpunct{(}{)}{;}{a}{}{,}
\let\oldthebibliography=\thebibliography
\let\endoldthebibliography=\endthebibliography
\renewenvironment{thebibliography}[1]{%
  \begin{oldthebibliography}{#1}%
    \setlength{\parskip}{0ex}%
    \setlength{\itemsep}{0ex}%
}%
{%
  \end{oldthebibliography}%
}

\end{document}